\documentclass[aps,prc,longbibliography,reprint,amssymb,amsmath,floatfix]{revtex4-1}

\usepackage[utf8]{inputenc}
\usepackage[T1]{fontenc}

\usepackage{newtxtext}
\usepackage{textcomp}
\usepackage{newtxmath}
\usepackage{bm}
\usepackage{amsmath}

\usepackage[english]{babel}
\usepackage{csquotes}

\usepackage{microtype}

\usepackage{graphicx}

\usepackage{float}
\usepackage{flafter}
\usepackage{pgf}
\usepackage{tikz}
\usepackage{color}

\usepackage{siunitx}
\usepackage{chemformula}
\setchemformula{
  subscript-style=math,
  charge-style=math,
  subscript-vshift=.0ex,
  charge-vshift=-.0ex,
  frac-style=nicefrac,
  plus-output-symbol=$+$,
}

\usepackage{xcolor}
\definecolor{DarkRed}{rgb}{0.65,0,0}
\definecolor{DarkBlue}{rgb}{0,0,0.65}
\definecolor{DarkGreen}{rgb}{0,0.65,0}

\usepackage[colorlinks=true,allcolors=blue]{hyperref}
\addto\extrasenglish{\def\equationautorefname~#1\null{Eq.~(#1)\null}}
\AtBeginDocument{

}

\renewcommand{\vec}[1]{\bm{#1}}
\providecommand{\drm}{\ensuremath{\mathrm{d}}}
\DeclareMathOperator{\real}{Re}
\DeclareMathOperator{\imag}{Im}
\DeclareMathOperator{\diag}{diag}
\providecommand{\up}{\uparrow}
\providecommand{\dn}{\downarrow}
\providecommand{\cre}[2]{#1^\dagger_{#2}}
\providecommand{\ann}[2]{#1^{\phantom{\dagger}}_{#2}}

\begin{document}

\title{Direct and inverse superspin Hall effect in two-dimensional systems: \\ Electrical detection of spin supercurrents}
\author{Vetle Risinggård}
\email{vetle.k.risinggard@ntnu.no}
\author{Jacob Linder}
\affiliation{Center for Quantum Spintronics, Department of Physics, NTNU, Norwegian University of Science and Technology, N-7491 Trondheim, Norway}
\date{\today}

\begin{abstract}
  A useful experimental signature of the ordinary spin Hall effect is 
  the spin accumulation it produces at the sample edges.
  The superspin Hall current [Phys.\ Rev.\ B~\textbf{96}, 094512 (2017)] is 
  a transverse equilibrium spin current which is induced by a charge supercurrent.
  We study the superspin Hall current numerically,
  and find that it does not give rise to a similar edge magnetization. 
  We also predict and numerically confirm the existence of 
  the inverse superspin Hall effect,
  which produces a transverse charge supercurrent in response to an equilibrium spin current.
  We verify the existence of the inverse superspin Hall effect both for
  a spin-polarized charge supercurrent and
  an exchange spin current,
  and propose that a $\phi_0$ junction produced by the inverse superspin Hall effect
  can be used to directly and electrically measure the spin polarization of a charge supercurrent.
  This provides a possible way to solve the long-standing problem of 
  how to directly detect the spin-polarization of supercurrents carried by triplet Cooper pairs.
\end{abstract}

\maketitle

\section{Introduction}
Spin-polarized supercurrents are a central theme in superconducting spintronics~\cite{
  Linder2015}. 
Cooper pairs in conventional Bardeen--Cooper--Schrieffer superconductors 
are in the spin-singlet state~\cite{
  Tinkham1996,
  DeGennes1999,
  Fossheim2004}.
Consequently, supercurrents in conventional superconductors are not spin polarized.
To spin polarize such a supercurrent,
the spin-singlet pairs must be converted to equal-spin triplet pairs.
This can be accomplished by combining the processes known as spin mixing and spin rotation~\cite{
  Linder2015,
  Eschrig2008,
  Eschrig2011,
  Eschrig2015}.
Because of the exchange splitting, 
proximity-induced Cooper pairs in a ferromagnet will oscillate between 
the spin-singlet and the spin-0 triplet state~\cite{
  Fulde1964,
  Larkin1965}.
This is known as spin mixing.
A magnetic inhomogeneity or spin--orbit coupling 
can rotate the resulting spin-0 triplets into equal-spin triplets~\cite{
  Bergeret2001,
  Bergeret2005,
  Buzdin2005,
  Bergeret2013,
  Bergeret2014}.
This is known as spin rotation.
So far, such a spin polarization of the supercurrent carried by triplet Cooper pairs
has not been detected directly,
but is only inferred from 
otherwise inexplicably long-ranged supercurrents in proximity structures~\cite{
  Eschrig2015}.

Long-ranged spin-polarized supercurrents in phase-biased Josephson junctions are 
equilibrium currents.
Various authors have suggested that spin supercurrents have 
observable consequences that can be detected via 
electrical~\cite{
  Mineev1992,
  Meier2003,
  Sonin2010} 
or mechanical~\cite{
  Sonin2007} means, 
or through the magnetization dynamics they induce~\cite{
  Kulagina2014,
  Bobkova2018}. 
Nonetheless, experimental detection schemes based on these signatures have yet to be implemented.
One particular difficulty with these suggestions is 
that an equilibrium spin current by definition cannot perform work without dissipating.
Consequently, any attempt to extract useful work from, 
say, a voltage induced by an equilibrium spin current in order to detect that current 
will dissipate the spin current itself.

The spin Hall effect~\cite{
  Sinova2015} and 
its Onsager reciprocal~\cite{
  Onsager1931,
  Onsager1931a,
  DeGroot1951} 
have found many applications in nonsuperconducting spintronics. 
Among others, these include 
electrical detection of spin currents induced by spin pumping~\cite{
  Saitoh2006} or 
the spin Seebeck effect~\cite{
  Uchida2008}; 
spin Hall magnetoresistance~\cite{
  Huang2012,
  Weiler2012};
and spin Hall spin-transfer torques~\cite{
  Ando2008}.
It is only natural to inquire whether 
a superconducting analog of the spin Hall effect can be used to detect 
the spin-polarization of a supercurrent.

Spin Hall effects in superconducting structures have been considered previously 
in several theoretical and experimental works.
Refs.~\cite{
  Kontani2009,
  Malshukov2011,
  Pandey2012,
  Malshukov2017,
  Espedal2017,
  Wakamura2015} 
considered out-of-equilibrium situations, 
in which quasiparticle effects give rise to spin (charge) currents as a result of 
charge (spin) injection. 
In particular, Ref.~\cite{
  Wakamura2015} 
measured an enhancement of the inverse spin Hall signal by 
three orders of magnitude when the NbN is cooled below the superconducting transition temperature. 
Refs.~\cite{
  Sengupta2006,
  Malshukov2010,
  Bergeret2016,
  Malshukov2008,
  Yang2012} 
considered equilibrium situations and 
it was shown that the combination of spin--orbit coupling and an exchange field could 
induce a phase difference between two superconductors to obtain a $\phi_0$-junction~\cite{
  Malshukov2010,
  Bergeret2016}.

\begin{figure}[b]
  \includegraphics[width=.9\columnwidth]{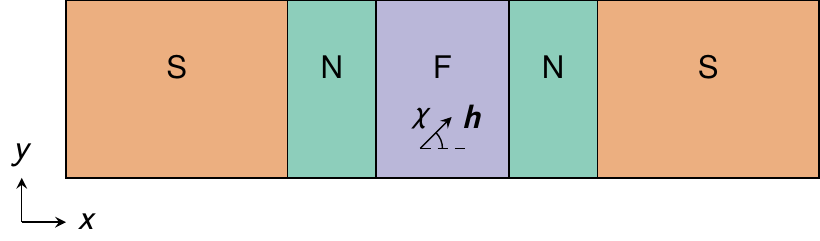}
  \caption{\label{fig:sysdirect}
  Suggested experimental setup for the superspin Hall effect. 
  A magnetic Josephson junction in the clean limit with Rashba spin--orbit interlayers. 
  The in-plane exchange field $\vec h$ in the ferromagnet makes an angle $\chi$ with the $x$ axis. 
  }
\end{figure}

In Ref.~\cite{
  Linder2017}, 
we considered an equilibrium transverse spin current generated by a 
longitudinal charge supercurrent in a Josephson junction, 
which we will refer to here as the \textit{superspin Hall current}.
Whereas most studies of spin Hall effects in superconductors consider 
purely $s$-wave or quasiparticle effects~\cite{
  Kontani2009,
  Malshukov2011,
  Pandey2012,
  Malshukov2017,
  Espedal2017,
  Wakamura2015,
  Sengupta2006,
  Malshukov2010,
  Bergeret2016,
  Malshukov2008},
the superspin Hall current is the result of an interplay between 
the $s$-wave condensate of a conventional superconductor and 
a proximity-induced $p$-wave condensate. 
As opposed to the interfacial spin current considered in Ref.~\cite{
  Yang2012}, 
the superspin Hall current considered in Ref.~\cite{
  Linder2017} 
arises in a magnetic Josephson junction. 
In Ref.~\cite{
  Linder2017} 
we also consider the ballistic limit, rather than
the diffusive limit considered in Refs.~\cite{
  Malshukov2008,
  Malshukov2010,
  Malshukov2011,
  Bergeret2016}. 

An open question regarding the superspin Hall current is whether or not 
it induces an edge spin magnetization 
which could serve as an experimental signature of its existence. 
This question was not addressed in Ref.~\cite{
  Linder2017}, 
which considered periodic boundary conditions and thus in practice a cylindrical geometry.

In this paper, we present two main results. 
The first result is a full two-dimensional analysis of the superspin Hall \textit{effect} 
where we are able to address the issue of what happens to the spin supercurrent 
at the edges of the system. 
This issue is of interest with respect to possible experimental probes of the effect. 

The second result is the prediction of a corresponding \textit{inverse effect,} 
namely the inverse superspin Hall effect. 
In this case, an equilibrium spin current produces a transverse charge supercurrent, 
which gives rise to a $\phi_0$ shift in the Josephson junction.
The $\phi_0$ shift is\ignorespaces
---as opposed to previous predictions of $\phi_0$ junctions incorporating
spin--orbit coupling and ferromagnets~\cite{
  Kulagina2014,
  Malshukov2010,
  Bergeret2016,
  Konschelle2015}---\ignorespaces
induced by a pure equilibrium spin current.
We propose that the $\phi_0$ shift can be used to detect 
the spin-polarization of a supercurrent carried by Cooper pairs.
Being an equilibrium property of the junction we consider, 
this detection scheme will not dissipate the equilibrium spin current. 
This offers a way to electrically and directly verify 
the spin polarization of previously detected long-ranged supercurrents~\cite{
  Keizer2006,
  Khaire2010,
  Robinson2010,
  Anwar2012,
  Martinez2016,
  Singh2016}.

The superspin Hall effect can not only be used to detect spin-polarized supercurrents,
but also other equilibrium spin currents.
To illustrate this we also calculate the $\phi_0$ shift induced by 
the exchange spin current between two misaligned ferromagnets~\cite{
  Slonczewski1989,
  Chen2014}.

\section{Introduction to the superspin Hall effect}
The intrinsic superspin Hall effect, 
which we considered in Ref.~\cite{
  Linder2017}, 
arises in a magnetic Josephson junction with Rashba spin--orbit interlayers, 
see \autoref{fig:sysdirect}. 
When a phase difference $\phi$ is applied over the junction, 
so that a longitudinal charge current flows between the two superconductors, 
a transverse spin current is induced near the superconductor--Rashba-metal interface. 
Being transverse, it flows parallel to the interface ($y$ direction). 
Its spin polarization is perpendicular to the exchange field $\vec h$ in the ferromagnet\ignorespaces
---along the $y$ direction for $\vec h=h\vec e_x$, 
and along the $x$ direction for $\vec h=h\vec e_y$. 

As we explain in Ref.~\cite{
  Linder2017}, 
this spin supercurrent is the result of a delicate interplay between 
the different condensates in the junction.
Consider for instance $\vec h=h\vec e_y$ and, 
for the sake of the argument, 
even-frequency superconducting correlations.
In addition to the $s$-wave spin-singlet condensate 
emanating from the proximitized superconductors, 
there are also $p$-wave correlations in the junction due to 
the broken translation symmetry at the material interfaces~\cite{
  Tanaka2007,
  Eschrig2007} and 
due to the presence of spin--orbit coupling~\cite{
  Sigrist1991}.
Due to the overall antisymmetry of the Cooper-pair wave function, 
the spin state of these even-frequency $p$-wave correlations must be one (or several) of 
the triplet states.
The generation of both short- and long-range triplets is possible because of 
the simultaneous presence of both ferromagnetism and spin--orbit coupling~\cite{
  Bergeret2013,
  Bergeret2014}. 
As explained in Ref.~\cite{
  Linder2017}, 
the interaction of the $s$- and $p$-wave condensates can be described via 
two different superconducting order parameters in the junction which 
quantify the superconducting correlations present in the system.
These are, respectively, 
the sum $\Delta_+$ and 
the difference $\Delta_-$ 
of the original $s$-wave and $p$-wave order parameters, $\Delta_s$ and $\Delta_k$, 
where $k$ refers to the momentum in the $y$ direction. 
The momentum index $k$ is a good quantum number for a system with 
periodic boundary conditions in the $y$ direction, 
as the one considered in Ref.~\cite{
  Linder2017}.
The relative magnitude of these order parameters is determined by 
the relative phase of the original $s$-wave and $p$-wave order parameters,
\begin{equation}\label{eq:sumdiff}
  |\Delta_\pm|^2=|\Delta_s|^2+|\Delta_k|^2\pm2\real(\Delta_s\Delta_k^*).
\end{equation}
When no phase difference is applied over the junction, 
the $s$-wave order parameter is purely real, 
whereas the $p$-wave order parameter is purely imaginary. 
Consequently, their sum and difference have equal magnitude, 
$|\Delta_+|=|\Delta_-|$, 
and as many Cooper pairs condense in 
the $|k\up,-k\dn\rangle$ state as in 
the $|k\dn,-k\up\rangle$ state. 
But, when a phase difference is applied, 
the $s$-wave order parameter acquires an imaginary component and 
the $p$-wave order parameter acquires a real component.
In turn, their sum and difference are no longer equal, 
$|\Delta_+|\neq|\Delta_-|$, 
and Cooper pairs condense preferentially at either 
$|k\up,-k\dn\rangle$ or 
$|k\dn,-k\up\rangle$ 
because of the difference in condensation energies. 
Such a selective condensation gives rise to 
a nonzero $k$-resolved spin magnetization $\vec S_k$ that is antisymmetric in $k$.
Subsequently, this antisymmetric momentum-resolved spin magnetization produces 
a spin current polarized along the spin magnetization direction.
For an exchange field $\vec h=h\vec e_y$ 
the momentum-resolved spin magnetization points in the $x$ direction; 
thus the application of a longitudinal phase difference (charge current) has given rise to 
a transverse spin current polarized along $\vec e_x$.

\section{Theory}
We consider a superconducting heterostructure in two dimensions in the clean limit, 
incorporating strong spin--orbit coupling. 
For this we use the tight-binding Bogoliubov--de~Gennes framework~\cite{
  DeGennes1999}. 
Our heterostructure consists of superconductors, 
normal metals with Rashba spin--orbit coupling, 
and ferromagnets.
Our Hamiltonian is
\begin{align}
  H=&-t\!\!\!\!\sum_{\langle i,j\rangle,\sigma}\!\!\!\cre{c}{i,\sigma}\ann{c}{j,\sigma}-\sum_{i,\sigma}\mu_i\cre{c}{i,\sigma}\ann{c}{i,\sigma}-\sum_iU_i\ann{n}{i,\up}\ann{n}{i,\dn} \notag\\
  &-(i/2)\!\!\!\!\!\sum_{\langle i,j\rangle,\alpha,\beta}\!\!\!\!\!\lambda_i[\vec n\cdot(\vec\sigma\times\vec d_{ij})]_{\alpha\beta}\cre{c}{i,\alpha}\ann{c}{j,\beta} \notag\\
  &+\sum_{i,\alpha,\beta}(\vec h_i\cdot\vec\sigma)_{\alpha\beta}\cre{c}{i,\alpha}\ann{c}{i,\beta}\,,
\end{align}
where $i$ and $j$ are position indices 
($i,j=1,\dots,N_xN_y$, where $N_x$ and $N_y$ are the dimensions of the lattice); 
$\langle i,j\rangle$ indicates that $i$ and $j$ are nearest neighbors;
$t$ is the hopping integral; 
$\cre{c}{i,\sigma}$ and $\ann{c}{i,\sigma}$ are electron creation and annihilation operators 
at site $i$ for spin $\sigma$; 
$\mu_i$ is the local chemical potential;
$U_i$ is the local on-site attraction that gives rise to superconductivity 
($U_i=0$ outside the superconductors and $U_i=U>0$ inside the superconductors);
$\ann{n}{i,\sigma}=\cre{c}{i,\sigma}\ann{c}{i,\sigma}$ is the number operator 
at site $i$ for spin $\sigma$;
$\lambda_i$ is the local Rashba parameter 
($\lambda_i=0$ outside the normal metals and $\lambda_i=\pm\lambda$ inside the normal metals);
$\vec n$ is the unit vector normal to the Rashba-metal/ferromagnet interface;
$\vec\sigma$ is the vector of Pauli matrices;
$\vec d_{ij}=-\vec d_{ji}$ is the vector pointing from site $i$ to site $j$;
and $\vec h_i$ is the local magnetic exchange field 
($\vec h_i=0$ outside the ferromagnet and $\vec h_i=\vec h$ inside the ferromagnet).

The two-particle Hubbard-$U$ term can be recast as
\begin{equation}\label{eq:hubbard}
  -\sum_iU_i\ann{n}{i,\up}\ann{n}{i,\dn}=\sum_i(\ann{\Delta}{i}\cre{c}{i,\dn}\cre{c}{i,\dn}+\cre{\Delta}{i}\ann{c}{i,\dn}\ann{c}{i,\up}+|\Delta_i|^2/U_i)
\end{equation}
using the standard mean-field \textit{ansatz} $\Delta_i=-U_i\langle c_{i,\dn}c_{i,\up}\rangle$.
We symmetrize the Hamiltonian using the fundamental fermionic anticommutator to write
\begin{equation}
  \sum_{\lambda,\kappa}A_{\lambda,\kappa}\cre{c}{\lambda}\ann{c}{\kappa}=\tfrac{1}{2}\sum_\lambda A_{\lambda,\lambda}+\tfrac{1}{2}\sum_{\lambda,\kappa}A_{\lambda,\kappa}(\cre{c}{\lambda}\ann{c}{\kappa}-\ann{c}{\kappa}\cre{c}{\lambda}).
\end{equation}
Introducing the basis
\begin{equation}
  \cre{B}{i}=\left(\begin{array}{cccc} \cre{c}{i,\up} & \cre{c}{i,\dn} & \ann{c}{i,\up} & \ann{c}{i,\dn} \end{array}\right),
\end{equation}
we may then write the Hamiltonian on the form
\begin{equation}\label{eq:elementform}
  H=H_0+\tfrac{1}{2}\sum_{i,j}\cre{B}{i}H_{ij}\ann{B}{j}.
\end{equation}
Here, we have identified the constant term $H_0$,
\begin{equation}
  H_0=\sum_i|\Delta_i|^2/U_i-\sum_i\mu_i,
\end{equation}
where the first sum runs only over the superconductors, and the $4\times4$ matrix $H_{ij}$,
\begin{align}
  H_{ij}=&\,\tfrac{1}{2}t\tau_z\sigma_0\delta_{j,i+\vec\delta}-\mu_i\tau_z\sigma_0\delta_{i,j}+\tfrac{i}{2}\ann{\Delta}{i}\tau_+\sigma_y\delta_{i,j} \notag\\
  &-\tfrac{i}{2}\cre{\Delta}{i}\tau_-\sigma_y\delta_{i,j}-\tfrac{i}{4}\lambda_i\tau_0\sigma_z(\delta_{j,i+\delta_y}-\delta_{j,i-\delta_y}) \notag\\
  &+h_i^x\tau_z\sigma_x\delta_{i,j}+h_i^y\tau_0\sigma_y\delta_{i,j}+h_i^z\tau_z\sigma_z\delta_{i,j}\,, \label{eq:Hij}
\end{align}
where $\delta_{i,j}$ is the Kronecker delta, 
we used $\vec n=\vec e_x$, 
we introduced the set of nearest neighbor vectors $\vec\delta=(\delta_x,\delta_y)$, and 
$\tau_n$ and $\sigma_n$ are the Pauli matrices for $n=x,y,z$ and $n=0$ refers to the identity. 
Moreover, $\tau_\pm=\tau_x\pm i\tau_y$, and 
products of Pauli matrices are interpreted as Kronecker products.
As is the usual definition, $\tau_z\sigma_0$, for instance, evaluates to 
$\tau_z\sigma_0=\diag(+1,+1,-1,-1)$~\cite{
  Dummit2004}. 

\begin{figure}[b]
  \includegraphics[width=.9\columnwidth]{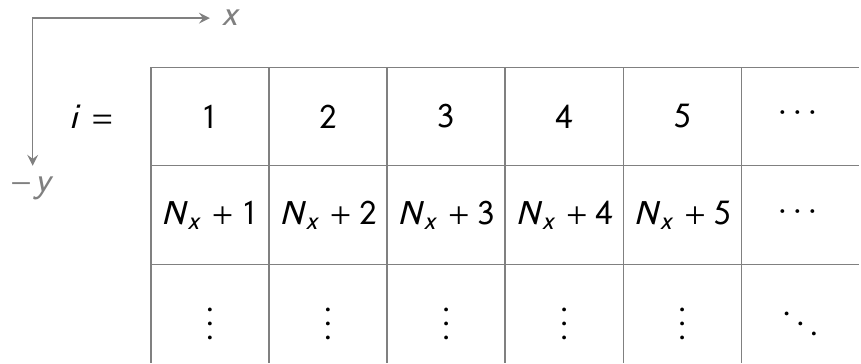}
  \caption{\label{fig:enumeration}
    Enumeration scheme for the $N_x\times N_y$ square lattice. 
    The site index $i$ is incremented site by site along the rows, starting in the upper left corner.
  }
\end{figure}

The index structure in \autoref{eq:elementform} is that of a matrix product, 
in which the matrix $M$ is multiplied from the left with the row vector $B^\dagger$, 
and the resulting row vector is multiplied with the column vector $B$. 
Each element in $M$ is a $4\times4$ matrix $H_{ij}$, 
and each element in $B$ (or $B^\dagger$) is a $4\times1$ (or $1\times4$) column (or row). 

The structure of the matrix $M$ is determined by 
how we combine the elements of $B$ and $B^\dagger$ into vectors.
We consider a square lattice.
The position indices $i$ and $j$ run over the entire system ($N_x\times N_y$ sites). 
Since each pair $(i,j)$ corresponds to a $4\times4$ block $H_{ij}$, 
we expect $M$ to be a $4N_xN_y\times4N_xN_y$ matrix. 

By choosing some enumeration scheme for the sites $i$ 
(such as the one in \autoref{fig:enumeration}), 
we can thus write
\begin{equation}\label{eq:matrixform}
  H=H_0+\tfrac{1}{2}B^\dagger MB,
\end{equation}
and diagonalize $M$ by the techniques that are familiar from linear algebra. 
Since $H$ is Hermitian, so is $M$, 
and $M$ can thus be diagonalized as $M=PEP^{-1}$, 
where $E$ is diagonal and real, and $P$ is unitary, $P^{-1}=P^\dagger$~\cite{
  Gantmacher2000}. 
Substituting $M=PEP^{-1}$ into \autoref{eq:matrixform} we obtain
\begin{equation}
  H=H_0+\tfrac{1}{2}\sum_nE_n\cre{\gamma}{n}\ann{\gamma}{n}\,,
\end{equation}
where we defined the new quasiparticle operators 
$\gamma^\dagger=B^\dagger P$ and 
$\gamma=P^{-1}B$, 
$\gamma_n$ is the $n$th element of $\gamma$, 
$E_n$ is the $n$th eigenenergy, 
and $n=1,\dots,4N_xN_y$.
The original electron operators can be related to the quasiparticle operators by
\begin{equation}
  \begin{split}
    &\ann{c}{i,\up}=\sum_nu_{i,n}\ann{\gamma}{n}\,, \qquad\ann{c}{i,\dn}=\sum_nv_{i,n}\ann{\gamma}{n}\,, \\
    &\cre{c}{i,\up}=\sum_nw_{i,n}\ann{\gamma}{n}\,, \qquad\cre{c}{i,\dn}=\sum_nx_{i,n}\ann{\gamma}{n}\,,
  \end{split}
\end{equation}
where $u_{i,n}$ with $i=1,\dots,N_xN_y$ is, respectively, $P_{ln}$ with $l=1,5,9,\dots$ 
Likewise, 
$v_{i,n}$ with $i=1,\dots,N_xN_y$ is $P_{ln}$ with $l=2,6,10,\dots$; 
$w_{i,n}$ with $i=1,\dots,N_xN_y$ is $P_{ln}$ with $l=3,7,11,\dots$; and 
$x_{i,n}$ with $i=1,\dots,N_xN_y$ is $P_{ln}$ with $l=4,8,12,\dots$. 

We can now derive expressions for any of the observables in the system 
in terms of the eigenenergies $E_n$ and 
the eigenvectors $u_{i,n}$, $v_{i,n}$, $w_{i,n}$, and $x_{i,n}$. 
For instance, the superconducting gap takes the form
\begin{equation}\label{eq:gap}
  \Delta_i=U_i\sum_nv_{i,n}w_{i,n}^*f(E_n/2),
\end{equation}
and the spin magnetization takes the form
\begin{subequations}
  \begin{align}
    &\langle S_i^x\rangle=2\sum_n\real(u_{i,n}^*v_{i,n})f(E_n/2), \label{eq:spinmagx} \\
    &\langle S_i^y\rangle=2\sum_n\imag(i_{i,n}^*v_{i,n})f(E_n/2), \\
    &\langle S_i^z\rangle=\sum_n(|u_{i,n}|^2-|v_{i,n}|^2)f(E_n/2).
  \end{align}
\end{subequations}
The free energy reads
\begin{equation}\label{eq:fenergy}
  F=H_0-\frac{1}{\beta}\sum_n\ln[1+\exp(-\beta E_n/2)],
\end{equation}
where $1/\beta=T$, and $T$ is temperature.

Expressions for the charge and spin currents can be obtained from 
their respective continuity equations,
\begin{equation}
  \partial_t\rho_i=-\nabla\cdot\vec j_i\,,
\end{equation}
and 
\begin{equation}
  \partial_t\vec s_i=-\nabla\cdot\vec J_i\,,
\end{equation}
where $\rho_i$ is the charge density at $i$, 
$\vec j_i$ is the current density at $i$, 
$\vec s_i$ is the spin density at $i$, 
$\vec J_i$ is the spin-current-density tensor at $i$, and 
the gradient of the spin-current-density tensor is taken with respect to the position variables. 
Note that the spin current defined by the spin continuity equation is only conserved in regions 
without ferromagnetism or spin--orbit coupling because 
these terms are spin nonconserving~\cite{
  Ouassou2017}.
For each of the two continuity equations, 
we find expressions for the currents by integrating the equations over space to obtain 
(for the case of the charge current)
\[
  \partial_tQ_i=-\int_\Omega\!\!\drm\vec r\,(\nabla\cdot\vec j_i),
\]
where $Q_i=\sum_\sigma\cre{c}{i,\sigma}\ann{c}{i,\sigma}$ is the charge at $i$ and 
$\Omega$ is the unit-cell volume.
The integral on the right-hand side can be evaluated using Green's theorem,
\[
  \int_\Omega\!\!\drm\vec r\,(\nabla\cdot\vec j_i)=\int_{\partial\Omega}\!\!\!\!\drm S\,(\vec j_i\cdot\vec e_n)=\sum_lj_{i,l}a=\sum_lI_{i,l},
\]
where $\partial\Omega$ is the unit-cell boundary, 
$\vec e_n$ is the outward-pointing boundary normal, and 
$a$ is the unit-cell side length. 
Since we consider a square lattice, 
$I_{i,l}$ is the current through the $l$th face of the square unit cell. 
The left-hand side of the continuity equation can be evaluated using 
Heisenberg's equation of motion. 
Thus the sum of currents out of the unit cell is
\begin{equation}
  \sum_lI_{i,l}=-i[H,Q_i].
\end{equation}
Evaluating the commutator and taking a combined thermal and quantum-mechanical average gives 
the charge current in the $x$ direction
\begin{align}
  &\langle I_i^x\rangle=t\sum_n\imag(u_{i+1,n}^*u_{i,n}-u_{i-1,n}^*u_{i,n} \notag\\
  &\hspace{5em}+v_{i+1,n}^*v_{i,n}-v_{i-1,n}^*v_{i,n})f(E_n/2)
\end{align}
and in the $y$ direction
\begin{align}
  \langle I_i^y\rangle=&\,t\sum_n\imag(u_{i-N_x,n}^*u_{i,n}-u_{i+N_x,n}^*u_{i,n} \notag\\
  &\hspace{4em}+v_{i-N_x,n}^*v_{i,n}-v_{i+N_x,n}^*v_{i,n})f(E_n/2) \notag\\
  &-\tfrac{1}{2}\sum_n\lambda_i\real(u_{i-N_x,n}^*u_{i,n}+u_{i+N_x,n}^*u_{i,n} \notag\\
  &\hspace{4em}-v_{i-N_x,n}^*v_{i,n}-v_{i+N_x}^*v_{i,n})f(E_n/2).
\end{align}
A similar procedure for the spin currents gives 
the three spin components of the spin current in the $x$ direction
\begin{align}
  &\langle I_i^{xx}\rangle=t\sum_n\imag\big(u_{i+1,n}^*v_{i,n}+v_{i+1,n}^*u_{i,n} \notag\\
  &\hspace{5em}-u_{i-1,n}^*v_{i,n}-v_{i-1,n}^*u_{i,n}\big)f(E_n/2), \\
  &\langle I_i^{xy}\rangle=t\sum_n\real\big(u_{i+1,n}^*v_{i,n}-v_{i+1,n}^*u_{i,n} \notag\\
  &\hspace{5em}-u_{i-1,n}^*v_{i,n}+v_{i-1,n}^*u_{i,n}\big)f(E_n/2), \\
  &\langle I_i^{xz}\rangle=t\sum_n\imag\big(u_{i+1,n}^*u_{i,n}-v_{i+1,n}^*v_{i,n} \notag\\
  &\hspace{5em}-u_{i-1,n}^*u_{i,n}+v_{i-1,n}^*v_{i,n}\big)f(E_n/2),
\end{align}
and likewise the three spin components of the spin current in the $y$ direction
\begin{align}
  &\langle I_i^{yx}\rangle=t\sum_n\imag\big(u_{i-N_x,n}^*v_{i,n}+v_{i-N_x,n}^*u_{i,n} \notag\\
  &\hspace{5em}-u_{i+N_x,n}^*v_{i,n}-v_{i+N_x,n}^*u_{i,n}\big)f(E_n/2), \\
  &\langle I_i^{yy}\rangle=t\sum_n\real\big(u_{i-N_x,n}^*v_{i,n}-v_{i-N_x,n}^*u_{i,n} \notag\\
  &\hspace{5em}-u_{i+N_x,n}^*v_{i,n}+v_{i+N_x,n}^*u_{i,n}\big)f(E_n/2), \\
  &\langle I_i^{yz}\rangle=t\sum_n\imag\big(u_{i-N_x,n}^*u_{i,n}-v_{i-N_x,n}^*v_{i,n} \notag \\
  &\hspace{5em}-u_{i+N_x,n}^*u_{i,n}+v_{i+N_x,n}^*v_{i,n}\big)f(E_n/2).
\end{align}

A general superconducting order parameter $F$ can be decomposed into 
a spin-singlet and a spin-triplet contribution~\cite{
  Leggett1975},
\begin{equation}
  F=(\psi+\vec d\cdot\vec\sigma)i\sigma_y\,,
\end{equation}
where $\psi$ is the singlet amplitude and 
the $\vec d$ vector is the vector of triplet amplitudes along the $x$, $y$, and $z$ axes,
\begin{equation}
  \vec d=\tfrac{1}{2}[\Delta_{\dn\dn}-\Delta_{\up\up},\,-i(\Delta_{\dn\dn}+\Delta_{\up\up}),\,2\Delta_{\up\dn}].
\end{equation}
The spin structure of the singlet amplitude is already familiar from \autoref{eq:Hij}, 
where the same factor $i\sigma_y$ appears.
In a unitary superconducting state, 
the identity $F^\dagger=F^{-1}$ holds, and $FF^\dagger$ is proportional to the identity. 
A general superconducting system is, however, not unitary, 
and straightforward calculation shows that
\begin{equation}
  FF^\dagger=|\psi|^2+|\vec d|^2+\vec\sigma\cdot[(\psi\vec d^*+\psi^*\vec d)+i(\vec d\times\vec d^*)].
\end{equation}
The term $i(\vec d\times\vec d^*)$ is proportional to 
the spin expectation value of the pure triplet Cooper pairs~\cite{
  Leggett1975}, 
whereas the term $(\psi\vec d^*+\psi^*\vec d)$ is proportional to 
the spin magnetization arising due to coexistence of singlet and triplet pairing~\cite{
  Linder2017},
\begin{equation}\label{eq:ampmag}
  \vec S_\text{Cooper}\propto(\psi\vec d^*+\psi^*\vec d)+i(\vec d\times\vec d^*).
\end{equation}
In order to calculate the Cooper-pair spin magnetization, 
we need expressions for the superconducting amplitudes.
The $s$-wave singlet amplitude $\mathcal S_{i,0}$ at $i$ is identical to 
the gap we calculated in \autoref{eq:gap}, 
except for the factor $U_i$,
\begin{equation}\label{eq:ss}
  \mathcal S_0=\tfrac{1}{2}[\langle\ann{c}{i,\up}\ann{c}{i,\dn}\rangle-\langle\ann{c}{i,\dn}\ann{c}{i,\up}\rangle]=\sum_nv_{i,n}w_{i,n}^*f(E_n/2).
\end{equation}
The direct and inverse superspin Hall effects depend on the existence of 
even-frequency, $p_y$-wave, spin-triplet amplitudes,
\begin{subequations}
  \begin{align}
    \mathcal P_{i,\up\dn}^y=&\,\tfrac{1}{2}\sum_\pm\pm[\langle\ann{c}{i,\up}\ann{c}{i\pm\delta_y,\dn}\rangle+\langle\ann{c}{i,\dn}\ann{c}{i\pm\delta_y,\up}\rangle] \notag\\
    =&\,\tfrac{1}{2}\sum_{n,\pm}\pm(w_{i,n}^*v_{i\mp N_x,n}-v_{i,n}w_{i\mp N_x,n}^*)f(E_n/2), \\
    \mathcal P_{i,\up\up}^y=&\sum_\pm\pm\langle\ann{c}{i,\up}\ann{c}{i\pm\delta_y,\up}\rangle \notag\\
    =&\sum_{n,\pm}\pm w_{i,n}^*u_{i\mp N_x,n}f(E_n/2), \\
    \mathcal P_{i,\dn\dn}^y=&\sum_\pm\pm\langle\ann{c}{i,\dn}\ann{c}{i\pm\delta_y,\dn}\rangle \notag\\
    =&\sum_{n,\pm}\pm x_{i,n}^*v_{i\mp N_x,n}f(E_n/2).
  \end{align}
\end{subequations}
In \autoref{sect:direct} we will need 
the odd-frequency, $s$-wave, spin-triplet amplitudes,
\begin{subequations}\label{eq:st}
  \begin{align}
    \mathcal S_{i,\up\dn}(t)=&\,\tfrac{1}{2}[\langle\ann{c}{i,\up}(t)\ann{c}{i,\dn}(0)\rangle+\langle\ann{c}{i,\dn}(t)\ann{c}{i,\up}(0)\rangle] \notag\\
    =&\,\tfrac{1}{2}\sum_n(w_{i,n}^*v_{i,n}-x_{i,n}^*u_{i,n})f(E_n/2)e^{iE_nt/2}, \\
    \mathcal S_{i,\up\up}(t)=&\,\langle\ann{c}{i,\up}(t)\ann{c}{i,\up}(0)\rangle \notag\\
    =&\sum_nw_{i,n}^*u_{i,n}f(E_n/2)e^{iE_nt/2}, \\
    \mathcal S_{i,\dn\dn}(t)=&\,\langle\ann{c}{i,\dn}(t)\ann{c}{i,\dn}(0)\rangle \notag\\
    =&\sum_nx_{i,n}^*v_{i,n}f(E_n/2)e^{iE_nt/2}.
  \end{align}
\end{subequations}

\section{Numerical calculations}
\begin{figure}[b]
  \includegraphics[width=.9\columnwidth]{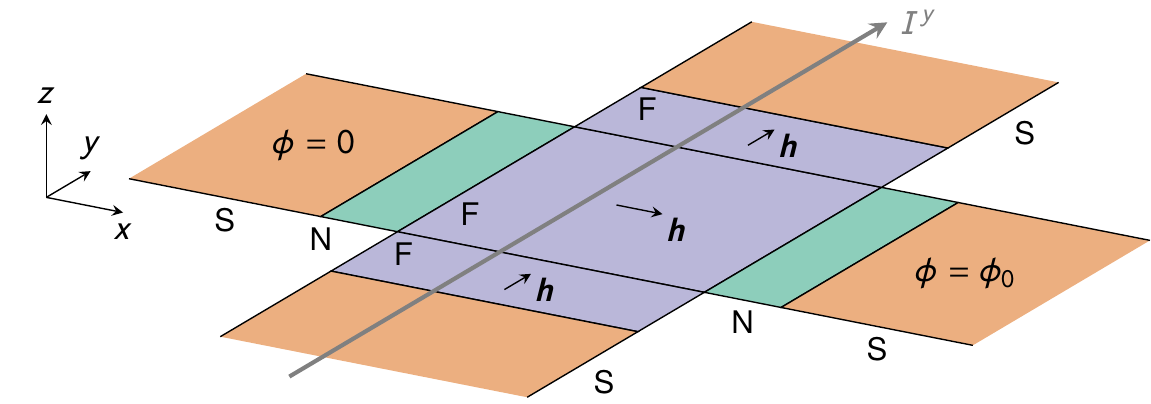}
  \caption{\label{fig:syscross}
  Proposed experimental setup for detecting a spin-polarized supercurrent 
  consisting of two crossed Josephson junctions. 
  The charge supercurrent in the $y$ direction injected into the S/F($y$)/F($x$)/F($y$)/S junction
  is spin polarized in the $x$ direction by 
  the magnetic inhomogeneity provided by the F($y$) layers.
  The transverse spin current thus injected into the S/N/F($x$)/N/S junction 
  induces a phase difference $\phi_0$ between the left and right superconductors. 
  }
\end{figure}

In this article, we consider the three setups in 
Figs.~\ref{fig:sysdirect}, \ref{fig:syscross}, and~\ref{fig:sysexchange}. 
In each case, we construct the matrix $M$ from \autoref{eq:matrixform} 
and diagonalize it to find the eigenvalues and eigenvectors.
Using these, we may calculate physical quantities such as 
the superconducting gap $\Delta_i$ or the spin magnetization $\langle\vec S_i\rangle$. 
Because the matrix $M$ depends on the superconducting gap, 
the equations must be solved self-consistently by 
substituting the gap calculated using \autoref{eq:gap} back into $M$ and iterating.

For each of the systems we consider, 
we make sure that the superconducting state minimizes the free energy in \autoref{eq:fenergy}.
In all the systems, 
we take the exchange field $\vec h_i$ of the ferromagnets to be an external parameter,
that is, we do not calculate the exchange field self-consistently.
This is consistent with an $s$--$d$-type model 
in which the localized $d$ electrons are responsible for the magnetic behavior~\cite{
  Coey2010,
  OHandley2000}.
The spin magnetization $\langle\vec S_i\rangle$ that we calculate is thus 
the spin polarization of the itinerant $s$ electrons.

In the setup in \autoref{fig:sysdirect} we consider 
the injection of a charge current into the junction by an applied phase difference.
This is accomplished by fixing the phase of the superconducting gap $\Delta_i$ 
at the leftmost lattice points in the left superconductor and 
at the rightmost lattice points in the right superconductor.
The applied phase difference between these points can be used as a proxy for 
the applied phase difference over the junction (N/F/N spacer) 
because the phase drop inside the superconductors is typically small.
(Fixing the phase difference at $\Delta\phi=0.5\pi$ gives 
an effective phase difference over the N/F/N spacer of $\Delta\phi\approx0.47\text{--}0.48\pi$.)

In the setup in \autoref{fig:syscross} we consider 
the injection of a charge current across the injection junction by an applied phase difference.
We fix the phase of the left superconductor in the detector at $\phi=0$ 
(this choice is arbitrary---only phase differences matter).
By varying the phase of the right superconductor from $0$ to $2\pi$
we calculate the current--phase and free-energy--phase relation of the detector junction.
We take the induced anomalous phase $\phi_0$ to be the phase over the detector
that minimizes the free energy and gives $\langle I^x\rangle=I(-\phi_0)=0$.

In the setup in \autoref{fig:sysexchange} we consider 
the injection of an exchange spin current from two misaligned ferromagnets.
We fix the phase of both superconductors at $\phi=0$ 
and calculate the anomalous charge current $I(0)=\langle I^x\rangle$.
In its simplest form, a $\phi_0$ junction~\cite{
  Liu2010,
  Liu2010a,
  Szombati2016,
  Rasmussen2016}
has the current--phase relation $I(\phi)=I_\text c\sin(\phi+\phi_0)$.
For small $\phi_0$ shifts, 
the anomalous phase $\phi_0$ and the anomalous current $I(0)=I_\text c\sin\phi_0$ are proportional, 
$I(0)\approx I_\text c\phi_0$.
Therefore, we can use the anomalous current as a proxy for the anomalous phase.

The advantage of tight-binding Bogoliubov--de~Gennes framework~\cite{
  DeGennes1999}
that we use
is that it is not subject to the limitations on length and energy scales 
that are inherent to for instance quasi-classical theory~\cite{
  Chandrasekhar2008}.
However, using this tight-binding framework,
only comparatively small lattice sizes are computationally manageable,
especially in two-dimensional finite-size calculations. 
For superconducting structures, 
the relevant length scale is the superconducting coherence length 
$\xi=\hbar v_\text F/\pi\Delta$~\cite{
  Tinkham1996,
  DeGennes1999}. 
If the coherence length is to be smaller than the thickness of the superconducting layers,
this requires relatively large values of the superconducting gap and large critical temperatures.
Nonetheless, the tight-binding framework can still be used to 
make qualitative and quantitative predictions for experimentally relevant systems. 
To do this requires that 
the spatial dimensions are scaled by the superconductive coherence length.
One example of a successful application of this method is Ref.~\cite{
  Black-Schaffer2010},
whose predictions correspond very well to the experimental results of Ref.~\cite{
  English2016}.

We take a similar approach.
With the parameters chosen in \autoref{sect:direct}--\ref{sect:exchange}
the thickness of the superconducting layers is about one coherence length,
and the normal-metal and ferromagnetic layers vary from about $\xi/4$ to $2\xi$.
As long as the weak links are not orders of magnitude larger than the coherence length,
the qualitative features of our results are robust towards variations of the system size.
In particular, the $\phi_0$ shift that we calculate in \autoref{sect:supercurrent}
is nearly independent of the length of the detector junction.

\section{Superspin Hall effect in two dimensions: \\ spin current and edge magnetization}
\label{sect:direct}
Our analysis of the superspin Hall effect in Ref.~\cite{
  Linder2017} 
was an effective one-dimensional analysis 
in the sense that we assumed periodic boundary conditions in the $y$ direction 
and thus could get rid of the $y$ coordinate by Fourier transformation. 
Whereas we were still able to calculate the transverse spin current, 
this left open the question of the exact spin current circulation pattern 
and whether any spin magnetization arises at the edges of the sample. 
The latter would be a useful experimental signature of the superspin Hall effect, 
as it has been previously for the (nonequilibrium) spin Hall effect~\cite{
  Kato2004,
  Wunderlich2005}. 

In the usual nonsuperconducting, nonequilibrium spin Hall effect, 
spin accumulates at the edges of the sample because 
the transverse spin current has nowhere to go upon reaching the sample boundary~\cite{
  Sinova2015}.
A steady state is achieved because 
the spin Hall effect is found in materials with strong spin--orbit coupling 
where spin is not conserved.
The accumulated spin at the edge at any time is thus the result of a balance between 
influx of spin from the bulk and spin loss due to spin--orbit coupling. 

\begin{figure}
  \includegraphics[width=\columnwidth]{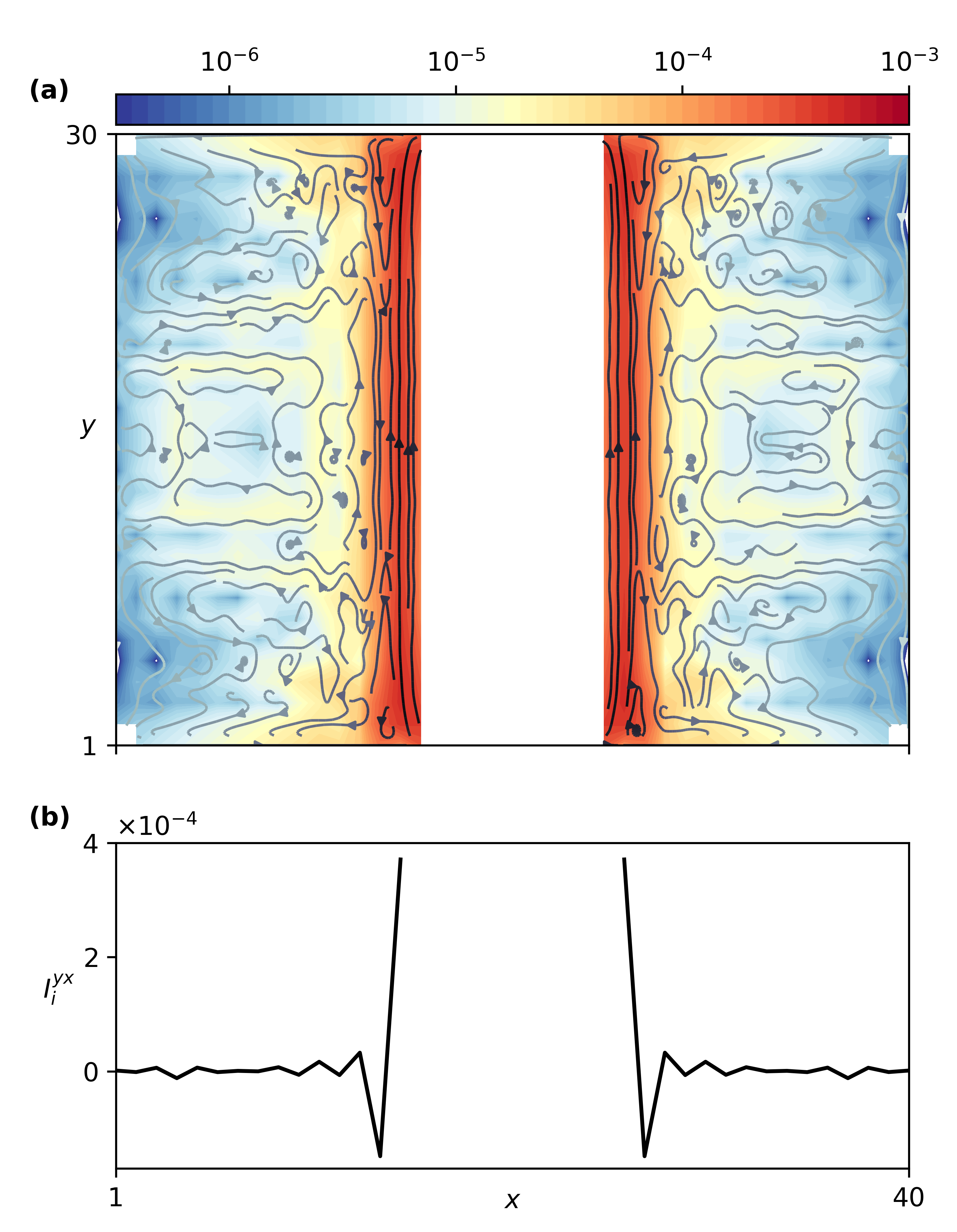}
  \vspace{-3ex}
  \caption{\label{fig:spincurrent}
  Superspin Hall current in two dimensions at phase difference $\phi=\pi/2$.
  (a) Circulation pattern of the $x$ component of the spin current.
  The spin current is only plotted in the superconductors, where spin is conserved. 
  (b) Cut along the $x$ direction at $y=15$ inside the superconductors.
  The spin current oscillates as a function of the distance from the N/F/N weak link into the superconductors. 
  We use the following parameter values:
  the system size is $N_x=40$ times $N_y=30$;
  the layer thicknesses are 
  $N_\text S=15$,
  $N_\text N=3$, and
  $N_\text F=4$.
  the chemical potentials are
  $\mu_\text S=0.9$,
  $\mu_\text N=0.85$, and
  $\mu_\text F=0.8$;
  the Rashba spin--orbit coupling in the normal metal is
  $\lambda=0.3$,
  the exchange field in the ferromagnet is
  $h_y=0.15$ ($h_x=h_z=0$),
  the on-site attraction in the superconductor is
  $U=1.1$,
  and the temperature is
  $T=0.01$.
  All energies are normalized with respect to the hopping parameter ($t=1$).
  }
\end{figure}

We find that the superspin Hall current does not give rise to a spin magnetization at the sample edges 
by this familiar mechanism.
The simple reason is that the superspin Hall current in our system does have somewhere to go---\ignorespaces
it can be drained from the superconductor, 
where in our model spin is conserved, 
into the Rashba-metal/ferromagnet spacer, 
where spin is not conserved. 
This circulation of the superspin Hall current from the spacer, 
into the superconductor, 
and back into the spacer, 
is shown in \autoref{fig:spincurrent}(a). 

Note that, although the net flow of spin is from the bottom of the sample to the top, 
the direction of the spin current (up/down) oscillates as a function of 
the distance into the superconductor~[\autoref{fig:spincurrent}(b)]. 
As explained in Ref.~\cite{
  Linder2017}, 
the oscillation period is a function of the system parameters, 
such as the strength of the spin--orbit coupling in the normal layer 
and the strength of the exchange field in the ferromagnet. 
The period varies from atomic-scale oscillations to roughly a fourth of the coherence length.
Such rapid oscillations are characteristic for 
physical quantities in ballistic quantum-mechanical systems.
For instance, they can also be found in 
the proximity-induced magnetization in conventional superconductors~\cite{
  Halterman2008} and 
helical edge-mode currents in triplet superconductors~\cite{
  Terrade2016}.

\begin{figure}
  \includegraphics[width=\columnwidth]{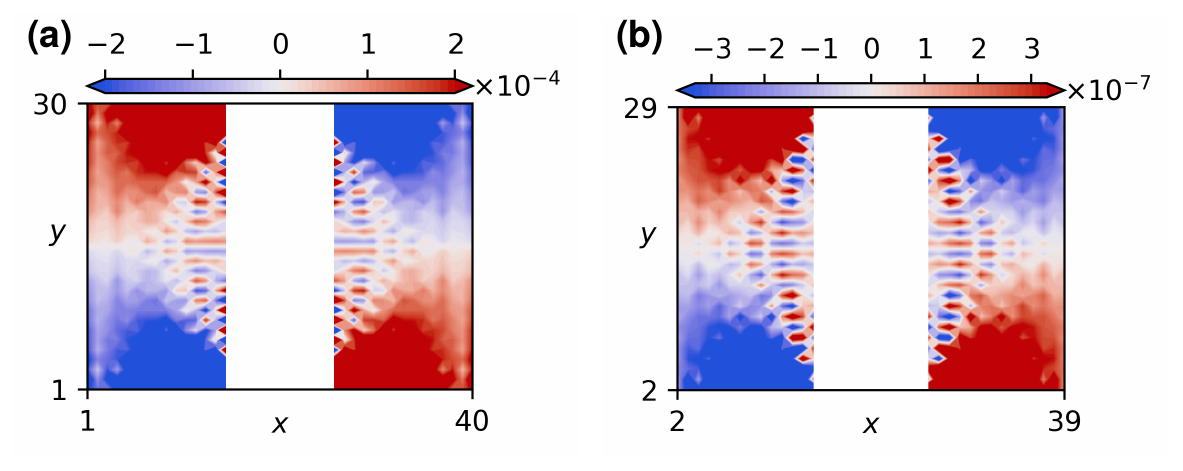}
  \caption{\label{fig:spinmag}
  $x$ component of the spin magnetization at phase difference $\phi=0$.
  (a) The total spin magnetization.
  (b) The spin magnetization induced by interaction of $s$-wave singlets and 
  odd-frequency, $s$-wave triplets (arbitrary units).
  Except for the applied phase between the superconductors, 
  all parameters are identical to \autoref{fig:spincurrent}.
  }
\end{figure}

\begin{figure}
  \hspace{-2cm}
  \includegraphics[width=.8\columnwidth]{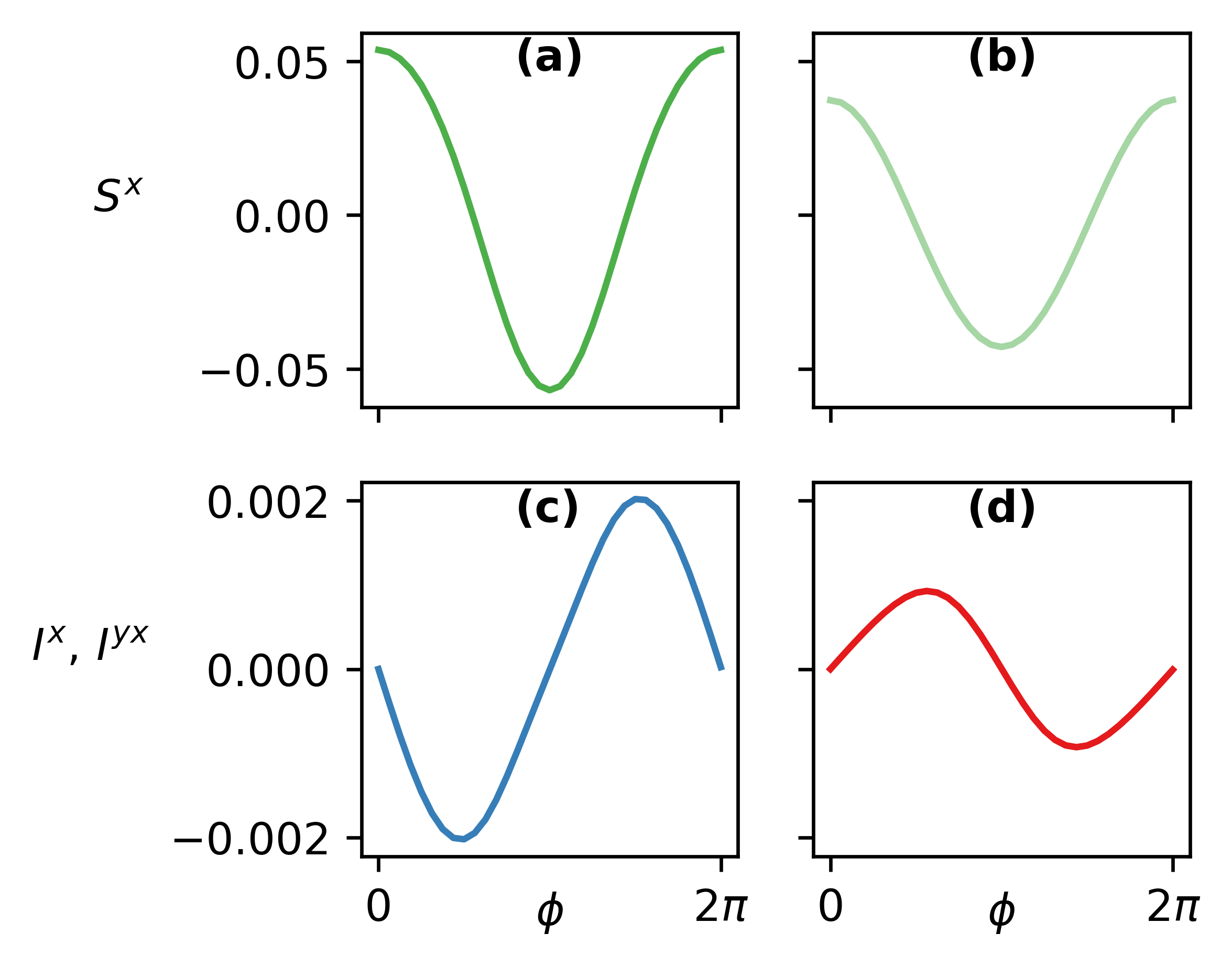}
  \caption{\label{fig:phidep}
  Phase dependence of the superspin Hall effect.
  (a)~Total spin magnetization $\langle S^x\rangle$ summed over the lower half of the right superconductor.
  (b)~Cooper-pair spin magnetization $S^x_\text{Cooper}$ summed over the lower half of the right superconductor (arbitrary units).
  (c)~Longitudinal charge current $\langle I^x\rangle$ summed over the $y$ cross section.
  (d)~Transverse spin current $\langle I^{yx}\rangle$ summed over the $x$ cross section.
  All parameters are identical to \autoref{fig:spincurrent}.
  }
\end{figure}

Although the superspin Hall current does not 
give rise to a spin magnetization at the edges of the sample, 
there \textit{is} an $x$-polarized spin magnetization at the edges of the system 
[\autoref{fig:spinmag}(a)]. 
However, contrary to what we would expect from 
a spin magnetization arising due to accumulation of spins deposited by the superspin Hall current,
the spin magnetization sign pattern that we observe is $\pm\mp$, not $\pm\pm$, 
where the signs refer to the left upper/lower and right upper/lower edges, respectively.
Furthermore, its amplitude varies as $\cos\phi$, 
where $\phi$ is the phase difference applied between the two superconductors 
[\autoref{fig:phidep}(a)].
We would expect a spin magnetization induced by the superspin Hall current\ignorespaces
---which is again induced by the longitudinal charge current---\ignorespaces
to have an amplitude that varied as $\sin\phi$ 
[compare with \autoref{fig:phidep}(c) and~(d)].

The momentum-resolved spin magnetization that gives rise to the superspin Hall current 
is the result of the interaction of the $s$-wave spin-singlet 
and a $p$-wave spin-triplet condensate,
both even in frequency.
The edge spin magnetization we observe in \autoref{fig:spinmag}, 
on the other hand, 
is the result of the interaction of the even-frequency, $s$-wave, spin singlet condensate 
and an odd-frequency, $s$-wave, spin-triplet condensate. 
In \autoref{fig:spinmag}(a) we have plotted 
the $x$ component of the total spin magnetization calculated using \autoref{eq:spinmagx}. 
In \autoref{fig:spinmag}(b) we have plotted 
the $x$ component of the spin magnetization calculated using \autoref{eq:ampmag}, 
where we have used the superconducting amplitudes in \autoref{eq:ss} and \autoref{eq:st}. 
Apart from a constant prefactor, the plots are essentially identical. 
The spin magnetization due to the odd-frequency, $s$-wave spin triplets 
also reproduce the phase dependence of the total spin magnetization 
[compare \autoref{fig:phidep}(a) and~(b)].

\begin{figure}
  \hspace{-1.5cm}
  \includegraphics[width=.8\columnwidth]{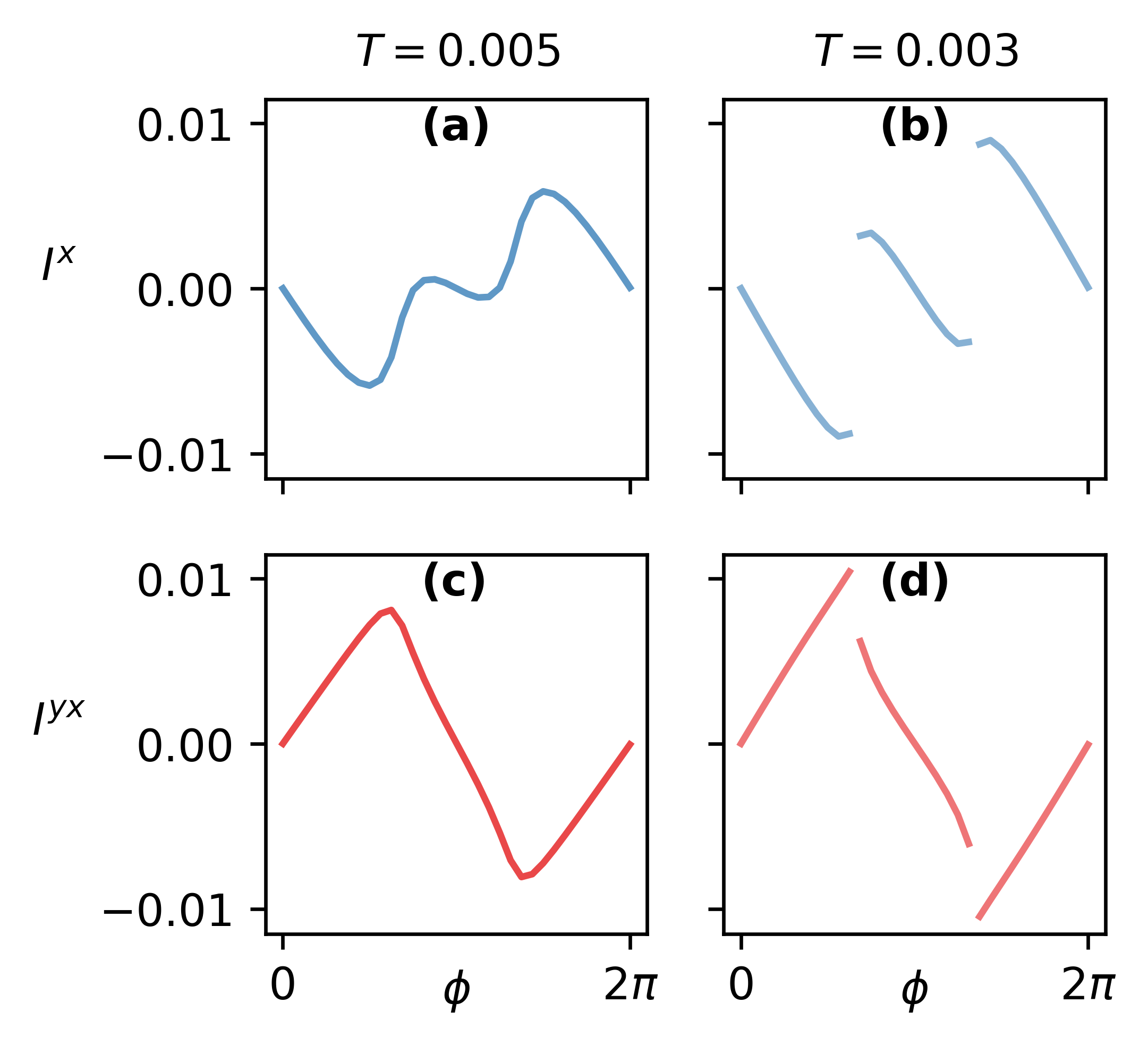}
  \caption{\label{fig:andreev}
  Phase dependence of the superspin Hall effect at low temperatures.
  (a)~Longitudinal charge current $\langle I^x\rangle$ at $T=0.005$ summed over the $y$ cross section.
  (b)~Longitudinal charge current $\langle I^x\rangle$ at $T=0.003$ summed over the $y$ cross section.
  (c)~Transverse spin current $\langle I^{yx}\rangle$  at $T=0.005$ summed over the $x$ cross section.
  (d)~Transverse spin current $\langle I^{yx}\rangle$  at $T=0.003$ summed over the $x$ cross section.
  All parameters except the temperature are identical to \autoref{fig:spincurrent}.
  }
\end{figure}

The fact that the edge spin magnetization is due to the odd-frequency triplets ($s$ wave) 
whereas the superspin Hall effect is due to the even-frequency triplets ($p$ wave) 
makes it clear that the spin magnetization is not a consequence of the superspin Hall current.
Further evidence to this effect 
is that this particular spin magnetization is also predicted in the diffusive limit~\cite{
  Hikino2018}, 
where the superspin Hall effect is precluded because of 
the absence of $p$-wave correlations. 
Consequently, one can exist independently of the other---they are independent effects. 

Nonetheless, 
the symmetries of the spin magnetization with respect to sign change of the Rashba spin--orbit coupling and 
the direction of the exchange field is the same as those of the superspin Hall current. 
In particular, rotating the exchange field by \SI{90}{\degree} 
from $\vec h=h\vec e_y$ to $\vec h=h\vec e_x$ 
also rotates the spin-triplet spin magnetization by \SI{90}{\degree} from $x$ to $y$. 

The temperature $T=0.01$ (in units of $t$), 
which we chose for the simulations above, 
is well below the superconducting transition temperature, $T=0.01\lesssim T_\text c/2$. 
However, at still lower temperatures, Andreev bound states~\cite{
  Andreev1964,
  Sauls2018} 
with a more dispersive energy--phase relation appear in the junction.
The appearance of such states is common in ballistic systems with 
high interface transparencies and low temperatures.
Because these states bounce multiple times between the two superconductors, 
they produce higher-harmonic contributions to the current--phase relation. 
The higher harmonics will distort the pure sinusoidal shape of the current--phase relation 
and may even introduce discontinuities~\cite{
  Kulik1977,
  Linder2008}. 
This, of course, also affects the superspin Hall current, as shown in \autoref{fig:andreev}. 

The presence of Andreev bound states in the junction also affects the spin magnetization, 
which deviates from a pure cosine as a function of the applied phase difference $\phi$. 
Interestingly, there is also a discernible difference between the total spin magnetization and 
the Cooper-pair spin magnetization computed via \autoref{eq:ampmag} at low temperatures.

\section{Inverse superspin Hall effect}
\label{sect:inverse}
The Onsager reciprocal of the usual nonsuperconducting, nonequilibrium spin Hall effect 
is the inverse spin Hall effect---\ignorespaces
that is, injection of a transverse spin current generates a longitudinal charge current.
In steady-state, the charge current must either be drained into external leads, 
or a voltage accumulates which
exactly cancels the inverse spin Hall current. 
Analogously, one might expect that there should exist an inverse of 
the superspin Hall effect discussed in \autoref{sect:direct}---\ignorespaces
injecting an equilibrium transverse spin current 
should give rise to a longitudinal charge supercurrent~\footnote{
  Note that the term \textit{inverse effect} cannot here be understood as 
  the Onsager reciprocal proper, 
  as our calculations are carried out is \textit{equilibrium}.}.
However, in the absence of external leads, 
the steady state will be one with zero charge current. 
Instead, a phase difference $\phi_0$ accumulates over the junction.
This phase difference gives rise to a supercurrent that 
exactly cancels the one induced by the inverse superspin Hall effect.
In this work, we confirm this expectation
and find that the experimental signature of the inverse superspin Hall effect 
is a $\phi_0$ junction.

\section{Electrical detection of \\ the supercurrent spin polarization}
\label{sect:supercurrent}
We propose to use the setup in \autoref{fig:syscross} to 
detect the spin polarization of a supercurrent.
This four-terminal setup consists of two perpendicular Josephson junctions. 
We will refer to 
the S/F($y$)/F($x$)/F($y$)/S junction as the injection junction and 
the S/N/F($x$)/N/S junction as the detector or detection junction. 

By applying a phase bias over the injection junction, 
a spin-polarized supercurrent is produced by the combined processes of 
spin mixing in the S/F($y$) bilayer 
and spin rotation (rotation of spin quantization axis between the F($y$) and F($x$) layers).
The current is spin-polarized in the $x$ direction.
The proximity to the F($y$) layers provides the necessary conditions for the superspin Hall mechanism.
Thus, the inverse superspin Hall effect converts this 
transverse spin current into a 
longitudinal charge supercurrent in the detector 
that flows from the left to the right superconductor. 
Consequently, in the steady state the detection junction is a $\phi_0$ junction.
If the two terminals of the detection junction are connected to form a superconducting loop, 
the current--phase relation of the detector can be measured by 
threading a magnetic flux through the loop~\cite{
  Golubov2004}. 
The anomalous current 
$I(0)=I_\text c\sin\phi_0$ 
can also be measured directly using a SQUID in zero applied flux.

\begin{figure}
  \hspace{-.5cm}
  \includegraphics[width=.8\columnwidth]{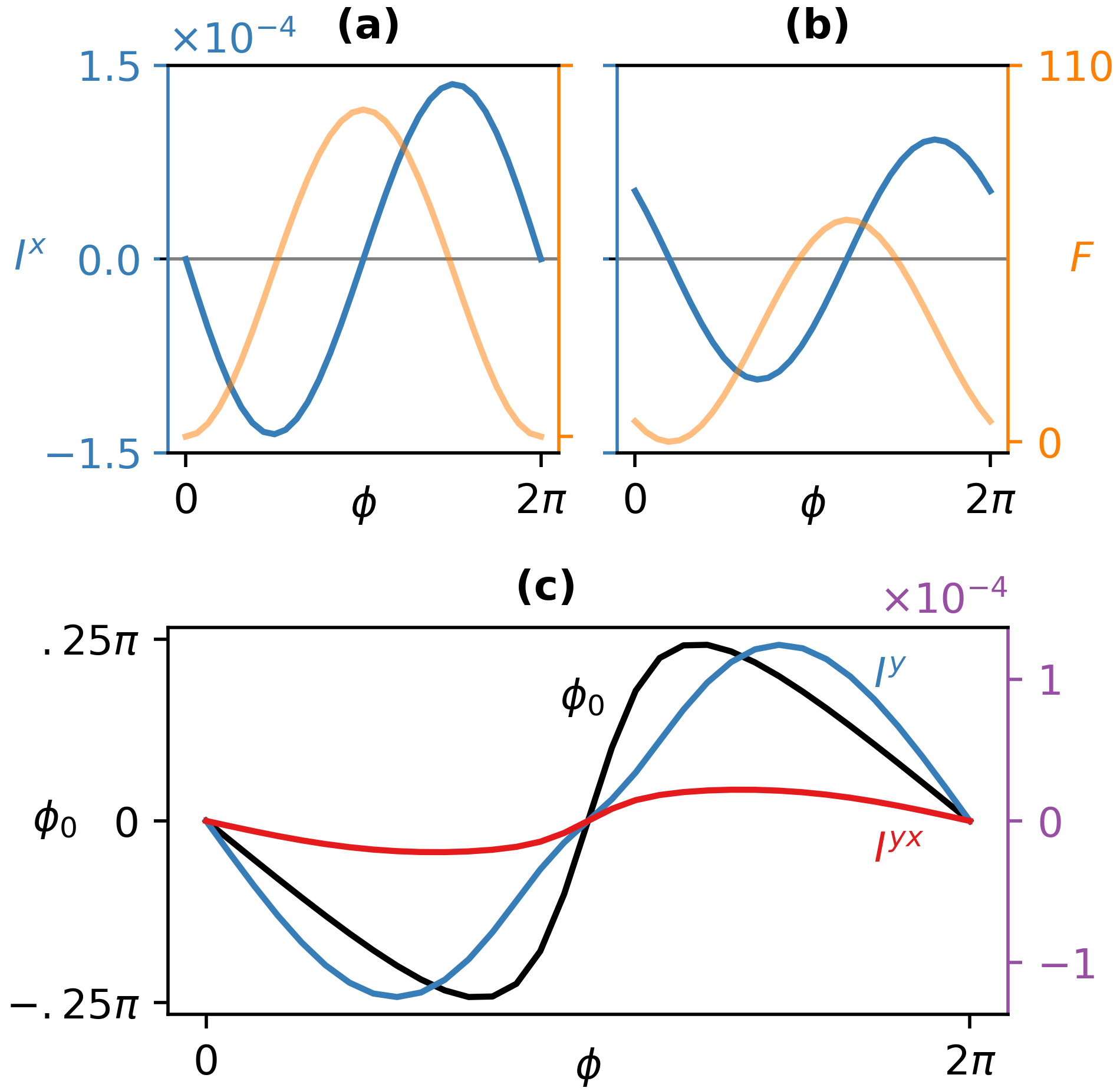}
  \caption{\label{fig:supercurrent}
  $\phi_0$ effect for the setup in \autoref{fig:syscross}. 
  (a)~and~(b): Current--phase relation and free-energy--phase relation of the detector junction
  as a function of the phase difference applied over the detector junction
  at applied phase differences of 
  (a)~$\phi=0$ (no injected charge current) and
  (b)~$\phi=\pi/2$ (maximal injected charge current)
  over the injection junction.
  In~(b) a $\phi_0$ shift of $\phi_0=-0.2\pi$ is clearly visible.
  (c)~Right ordinate:
  the injected charge current and 
  the resulting spin current 
  through the injection junction 
  as a function of the phase difference applied over the injection junction.
  Left ordinate: 
  the induced $\phi_0$ shift in the detection junction
  as a function of the phase difference applied over the injection junction.
  The system size is $N_x=35$ times $N_y=21$.
  The layer thicknesses of the detector are 
  $N_\text S=5$,
  $N_\text N=10$, and
  $N_\text F=5$,
  and the layer thicknesses of the injector are
  $N_\text S=5$,
  $N_{\text F(x)}=3$, and
  $N_{\text F(y)}=5$.
  All material parameters are identical to \autoref{fig:spincurrent}.
  }
\end{figure}

\autoref{fig:supercurrent}(a) and~(b) show 
the current--phase and 
the free-energy--phase relation 
of the detector junction. 
At an applied phase difference of $\phi=0$ over the injector junction 
[\autoref{fig:supercurrent}(a)] 
no spin current is injected across the detector. 
Consequently, the current--phase relation of the detector junction is that 
of an ordinary 0~junction. 
At an applied phase difference of $\phi=\pi/2$ over the injector junction 
[\autoref{fig:supercurrent}(b)] 
a large spin current is injected across the detector. 
Consequently, the current--phase relation is shifted by an amount $\phi_0=-0.2\pi$. 

\autoref{fig:supercurrent}(c) shows the complete $\phi_0$--phase relation. 
The abscissa corresponds to the applied phase difference of the injection junction. 
On the right ordinate we have plotted the charge and spin currents injected across the detector, 
that is, $\langle I^y\rangle$ and $\langle I^{yx}\rangle$. 
The $\langle I^y\rangle$--phase and the $\langle I^{yx}\rangle$--phase relations 
are both almost sinusoidal, 
and we interpret the spin current as the spin polarization of the charge current. 
On the left ordinate we have plotted the induced $\phi_0$ shift, 
\textit{i.e.}\ the phase $\phi$ over the detector that corresponds to 
$\langle I^x\rangle=0$ and $F=F_\text{min}$. 
Clearly, the $\phi_0$ shift is zero when the transverse spin current is zero. 
Moreover, the sign of the $\phi_0$ shift is a good predictor for the sign of the spin current. 
We have not been able to find a simple explanation for 
the deviation of the $\phi_0$ shift from a pure sine, 
but the fact that 
both the sign and zeros of the anomalous phase $\phi_0$ follow the spin supercurrent 
is consistent with the latter being the origin of the anomalous phase shift.

In addition to serving as a measurement of the spin polarization of the supercurrent, 
the setup we propose in \autoref{fig:syscross} can also serve as a current-controlled phase battery.
Such functionality has recently been proposed for a voltage-controlled $\phi_0$ junction~\cite{
  Ouassou2018}, 
and recent experiments have made progress towards both magnetic and electric phase control~\cite{
  Szombati2016,
  Glick2018}.

\begin{figure}
  \includegraphics[width=.9\columnwidth]{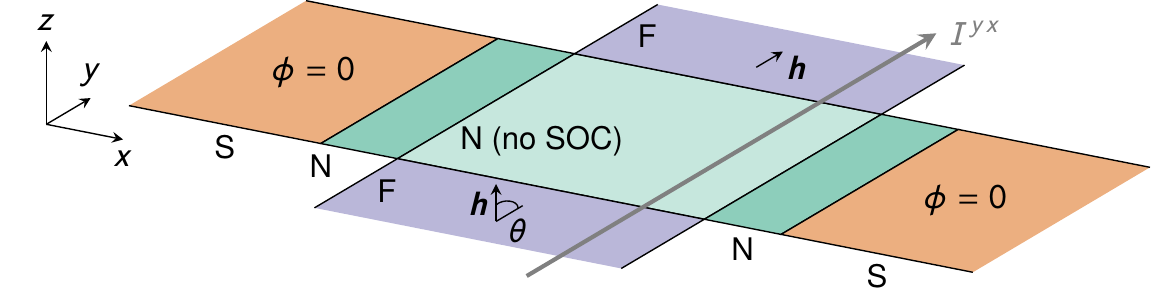}
  \caption{\label{fig:sysexchange}
  Suggested experimental setup for the inverse superspin Hall effect.
  The misalignment of the two ferromagnets 
  (misalignment angle $\theta$)
  produces a transverse exchange spin current
  that gives rise to an anomalous current between the two superconductors. 
  }
\end{figure}

\begin{figure}[b]
  \hspace{-.5cm}
  \includegraphics[width=.8\columnwidth]{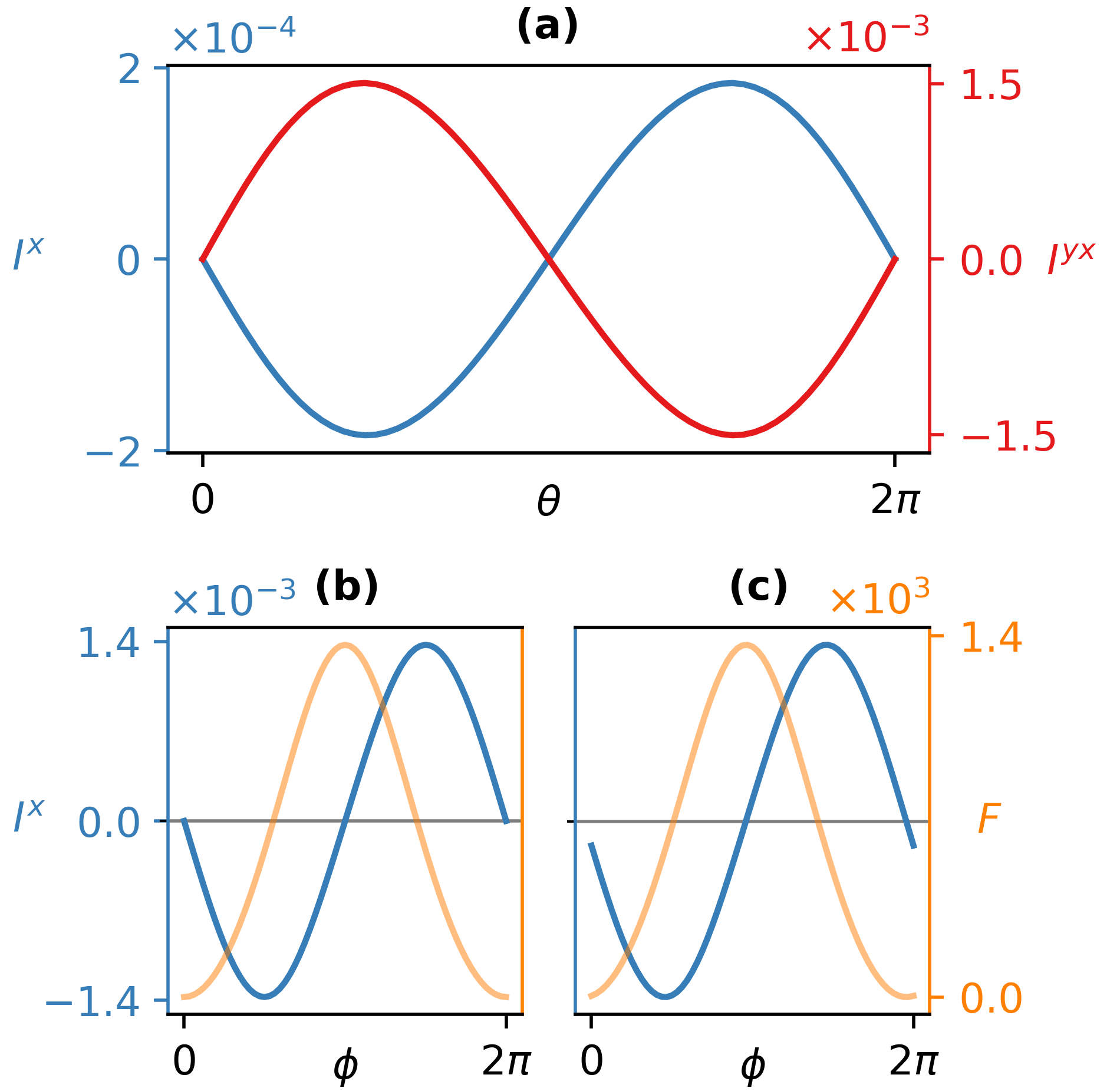}
  \caption{\label{fig:exchange}
  (a)~Injected spin current $\langle I^{yx}\rangle$ in the central normal metal
  (no spin--orbit coupling) and
  anomalous current $I(0)=\langle I^x\rangle$ 
  as a function of the misalignment angle $\theta$.
  (b)~and (c): Current--phase relation and 
  free-energy--phase relation 
  at a misalignment angle of respectively 
  $\theta=0$ (no injected spin current) and
  $\theta=\pi/2$ (maximal injected spin current). 
  A $\phi_0$ shift of $\phi_0\approx-0.04\pi$ is clearly visible. 
  We use the following parameter values:
  the system size is $N_x=38$ times $N_y=12$;
  the layer thicknesses of the detector are
  $N_\text S=15$,
  $N_\text N=2$, and
  $N_{\text N'}=4$
  and the layer thicknesses of the injector are
  $N_\text F=1$ and
  $N_{\text N'}=10$;
  the Rashba spin--orbit coupling is
  $\lambda=1.87$; and
  the exchange field is
  $h=0.8$.
  The remaining parameter values are identical to \autoref{fig:spincurrent}.
  }
\end{figure}

\section{Electrical detection of \\ an exchange spin current}
\label{sect:exchange}
The inverse superspin Hall effect is not only induced by
a spin-polarized charge supercurrent,
but also by other equilibrium spin currents.
To demonstrate this, we consider the setup in \autoref{fig:sysexchange}. 
Here, the injection junction has been replaced by an F/N/F spin valve 
(no spin--orbit coupling in N).
By misaligning the ferromagnets, we can inject an exchange spin current~\cite{
  Slonczewski1989,
  Chen2014}. 
The spin current is proportional to the sine of the misalignment angle $\theta$, 
$I_\text s\sim\sin\theta$. 

In \autoref{fig:exchange}(a) we plot 
the resulting spin current $\langle I^{yx}\rangle$ in the central normal metal
(where spin is conserved) and
the anomalous current $I(0)=\langle I^x\rangle$ 
as a function of the misalignment angle $\theta$. 
The spin current is sinusoidal as a function of $\theta$, 
consistent with the prediction of Ref.~\cite{
  Slonczewski1989,
  Chen2014}.
The sinusoidal response of the anomalous current 
is consistent with our interpretation that it is induced by the exchange spin current
(and not directly induced by the transverse variation in the exchange field as in Ref.~\cite{
  Malshukov2010}).

In \autoref{fig:exchange}(b) and~(c) we plot the current--phase relation 
at a misalignment angle $\theta=0$ and $\theta=\pi/2$. 
The anomalous current $I(0)$ shows up as a $\phi_0$ shift of the current--phase relation.

\begin{figure}
  \hspace{-1cm}
  \includegraphics[width=.9\columnwidth]{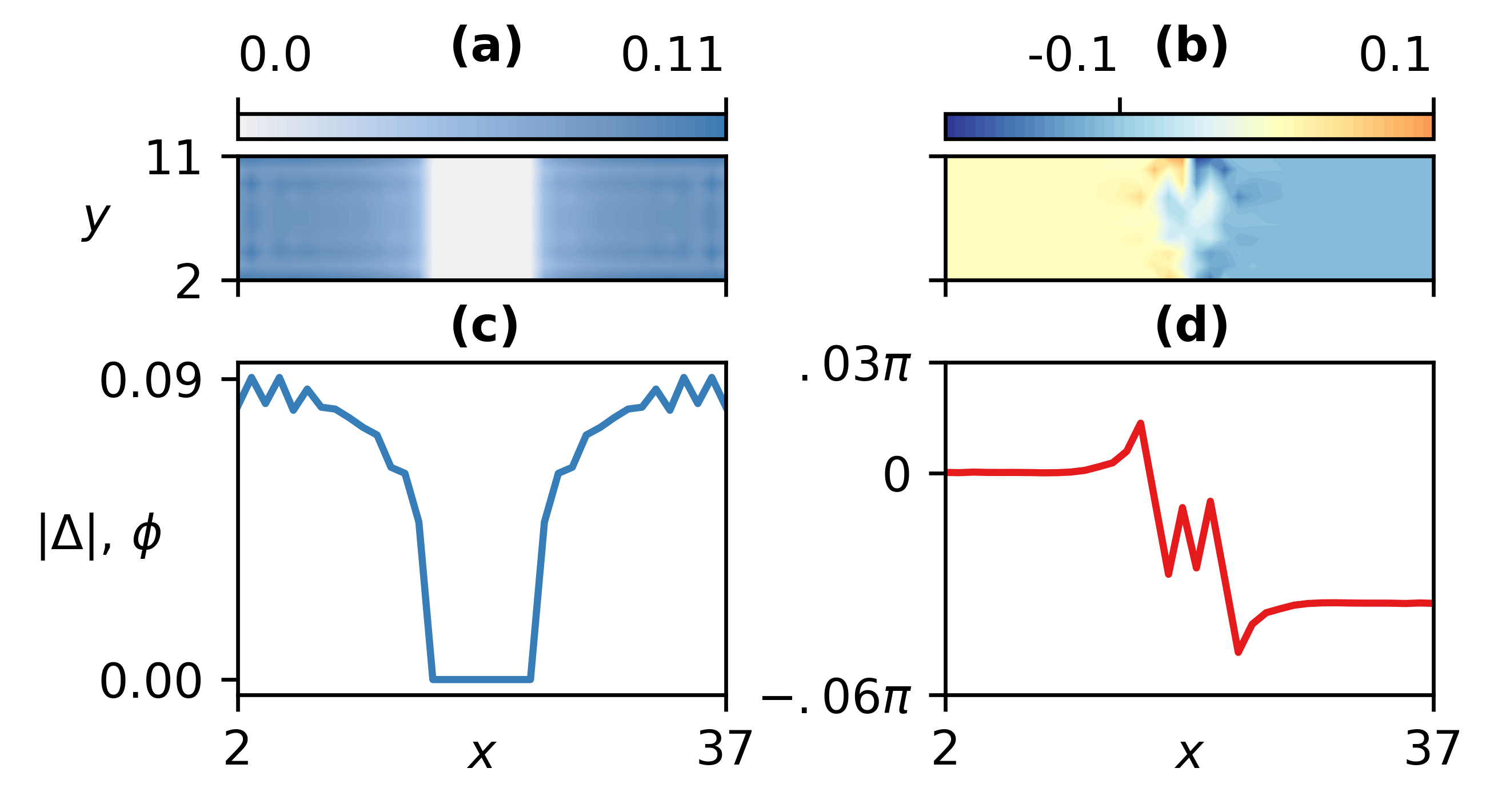}
  \caption{\label{fig:gapexchange}
  $\phi_0$ signature of the inverse superspin Hall effect.
  (a)~and (c): Magnitude of the superconducting gap $|\Delta_i|$. 
  The gap vanishes in the N/F/N spacer and the inverse proximity effect is clearly visible. 
  The oscillations of the gap at the edges of the sample are due to Fridel oscillations.
  (b)~and (d): Phase of the $s$-wave singlet amplitude $\mathcal S_{i,0}$.
  A $\phi_0$ shift of $\phi_0\approx-0.04\pi$ is clearly visible.
  The plots in panels~(c) and~(d) are for $y=5$.
  We use  parameter values that are identical to \autoref{fig:exchange}.
  }
\end{figure}

\begin{figure}[b]
  \hspace{.3cm}
  \centering
  \includegraphics[width=.8\columnwidth]{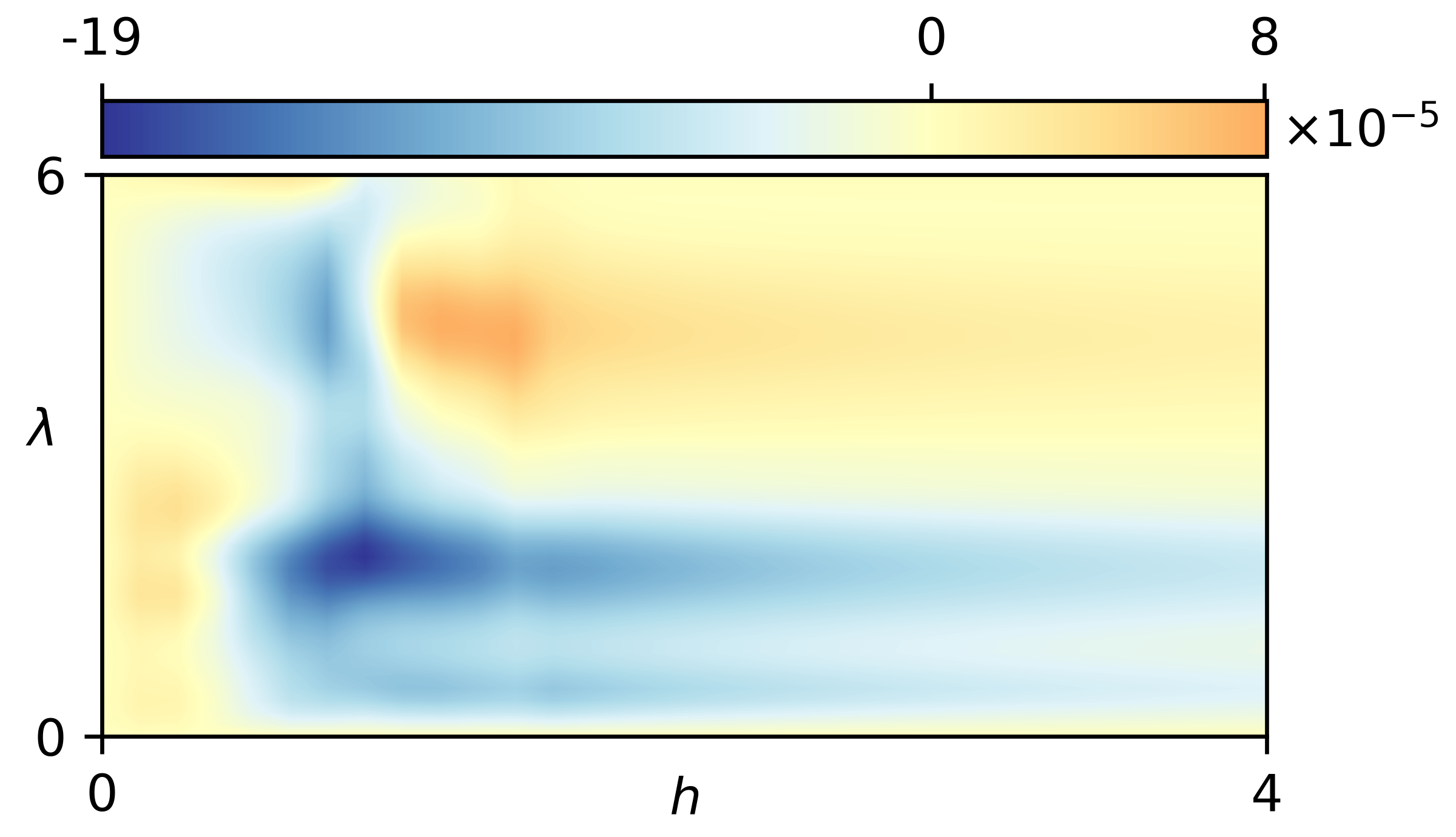}
  \caption{\label{fig:params}
  Dependence of the anomalous current $I(0)=\langle I^x\rangle$ on 
  the ferromagnet exchange field $h$ and 
  the Rashba metal spin--orbit coupling $\lambda$.
  Except for $h$ and $\lambda$,
  the parameter values are identical to \autoref{fig:exchange}.
  }
\end{figure}

In \autoref{fig:gapexchange} we have plotted 
the magnitude and phase of the resulting superconducting gap.
The oscillations in the gap magnitude $|\Delta_i|$ at the sample edges 
are due to Fridel oscillations
that create an oscillating change density~\cite{
  Ashcroft1976}.
A $\phi_0$ shift of $\phi_0\approx-0.04\pi$ is clearly visible.

\autoref{fig:params} shows the dependence of the anomalous current $I(0)$ on 
the spin--orbit coupling strength $\lambda$ in the Rashba metals and 
the exchange-field strength $h$ in the ferromagnets. 
There is a pronounced peak (or dip) at $h\approx0.8$ and $\lambda\approx1.9$. 
This parameter dependence can be understood as follows:
We expect the inverse superspin Hall effect to disappear when the exchange field vanishes
because $h=0$ means that no transverse spin current is injected.
(Also, magnetism is a prerequisite for the superspin Hall effect.)
For small values of $h$, we expect the anomalous current to increase with the exchange field
because an increase in $h$ leads to an increase in the transverse spin current.
However, for large values of $h$ we expect the superspin Hall effect to disappear 
because the exchange field suppresses the superconducting proximity effect.

We also expect the anomalous current to vanish for vanishing spin--orbit coupling
because spin--orbit coupling is a prerequisite for the superspin Hall effect.
For finite $\lambda$ there is a finite anomalous current
because of the superspin Hall effect,
but we expect the superspin Hall effect to disappear for very large spin--orbit coupling
because it suppresses the necessary $p_y$-wave spin-$0$ triplets \cite{
  Linder2017}
($\vec d$ not parallel to $\vec g_{\vec k}$ in the notation of Ref.~\cite{
  Frigeri2004}).

\section{Discussion}
The superspin Hall effect and its inverse depend on 
the existence of $p$-wave correlations in the junction.
These correlations are sensitive to disorder and will,
in the face of too large amounts of disorder,
be entirely suppressed. 

The suppression of superconductivity by disorder has been studied in many systems,
including 
heavy-fermion systems~\cite{
  Dalichaouch1995,
  Geibel1994},
iron pnictides~\cite{
  Onari2009,
  Wang2013},
and 
\ch{Sr2RuO4}~\cite{
  Mackenzie1998}.
Strontium ruthenate is arguably the most relevant system for 
the $p$-wave correlations that the superspin Hall effect depends on.
In strontium ruthenate, 
the disorder-dependence of the critical temperature can be described using
Abrikosov--Gor'kov pair-breaking theory~\cite{
  Abrikosov1961,
  DeGennes1999}.
Superconductivity vanishes in this compound when 
the mean-free path $\ell$ is on the order of or smaller than 
the superconducting coherence length $\xi$ of the $p$-wave order parameter.
Experiments indicate that this corresponds to a residual resistivity of about 
\SI{1}{\micro\ohm\centi\meter}~\cite{
  Mackenzie1998}.
Results from the iron pnictides indicates that $s_\pm$-wave pairing is suppressed
at a similar residual resistivity of about \SI{10}{\micro\ohm\centi\meter}~\cite{
  Wang2013},
corresponding to an impurity concentration of only about \SI{1}{\percent}~\cite{
  Onari2009}.

We expect that a similar strong suppression of the $p$-wave correlations
will take place in the junctions we consider.
To realize the effects we predict experimentally would thus require 
samples with good crystallinity and impurity concentrations below about \SI{1}{\percent}.

In the weakly disordered case---\ignorespaces
that is, for impurity concentrations below this level---\ignorespaces
we expect that the amount of $p$-wave correlations will be reduced, 
but not have vanished completely.
This will lead to a reduction in the induced transverse spin current
(superspin Hall effect)
or the induced anomalous current
(inverse superspin Hall effect)
compared to the clean limit. 
For comparison, it is instructive to compare the behavior in this case to \autoref{fig:params}.
Here, the $p_y$-wave spin-0 triplets are suppressed at large spin--orbit coupling,
and the anomalous current vanishes. 
Similar behavior can be expected as a function of impurity concentration.

\vspace{1cm}

\section{Conclusion}
We have considered the superspin Hall and the inverse superspin Hall effects 
in a two-dimensional S/N/F/N/S Josephson junction.
We present two main results.

Firstly, the transverse spin supercurrent induced by the superspin Hall effect 
circulates from the N/F/N spacer, into the superconductors, and back into the N/F/N spacer. 
Consequently, it does not give rise to a spin magnetization at the sample edges, 
contrary to the usual spin Hall effect. 
The spin magnetization that \textit{does} arise at the sample edges 
can be attributed to interaction between 
the proximity-induced even-frequency $s$-wave spin-singlet condensate and 
odd-frequency $s$-wave spin-triplet correlations.

Secondly, we predict and numerically confirm the existence of the inverse superspin Hall effect,
which can be detected experimentally as a $\phi_0$ shift in the Josephson junction.
We have shown that both exchange spin currents and 
spin-polarized charge supercurrents
produce a transverse charge supercurrent by the inverse superspin Hall effect. 
In particular, we propose that a $\phi_0$ junction produced by
the inverse superspin Hall effect 
can be used to\ignorespaces
---for the first time---\ignorespaces
measure directly the spin polarization of a charge supercurrent
carried by triplet Cooper pairs.

\begin{acknowledgments}
  We would like to thank M.\ Amundsen and J.\,A.\ Ouassou for useful discussions.
  Funding via the ``Outstanding Academic Fellows'' program at NTNU, 
  the NV Faculty, 
  the Research Council of Norway Grant No.\ 240806, 
  and the Research Council of Norway through its Centres of Excellence funding scheme, 
  Project No.\ 262633, ``QuSpin'' is gratefully acknowledged. 
  This research has benefited from the Notur high-performance computing facilities, 
  Project No.\ NN9577K.
\end{acknowledgments}

\vspace{2cm}


\begin{thebibliography}{87}%
\makeatletter
\providecommand \@ifxundefined [1]{%
 \@ifx{#1\undefined}
}%
\providecommand \@ifnum [1]{%
 \ifnum #1\expandafter \@firstoftwo
 \else \expandafter \@secondoftwo
 \fi
}%
\providecommand \@ifx [1]{%
 \ifx #1\expandafter \@firstoftwo
 \else \expandafter \@secondoftwo
 \fi
}%
\providecommand \natexlab [1]{#1}%
\providecommand \enquote  [1]{``#1''}%
\providecommand \bibnamefont  [1]{#1}%
\providecommand \bibfnamefont [1]{#1}%
\providecommand \citenamefont [1]{#1}%
\providecommand \href@noop [0]{\@secondoftwo}%
\providecommand \href [0]{\begingroup \@sanitize@url \@href}%
\providecommand \@href[1]{\@@startlink{#1}\@@href}%
\providecommand \@@href[1]{\endgroup#1\@@endlink}%
\providecommand \@sanitize@url [0]{\catcode `\\12\catcode `\$12\catcode
  `\&12\catcode `\#12\catcode `\^12\catcode `\_12\catcode `\%12\relax}%
\providecommand \@@startlink[1]{}%
\providecommand \@@endlink[0]{}%
\providecommand \url  [0]{\begingroup\@sanitize@url \@url }%
\providecommand \@url [1]{\endgroup\@href {#1}{\urlprefix }}%
\providecommand \urlprefix  [0]{URL }%
\providecommand \Eprint [0]{\href }%
\providecommand \doibase [0]{http://dx.doi.org/}%
\providecommand \selectlanguage [0]{\@gobble}%
\providecommand \bibinfo  [0]{\@secondoftwo}%
\providecommand \bibfield  [0]{\@secondoftwo}%
\providecommand \translation [1]{[#1]}%
\providecommand \BibitemOpen [0]{}%
\providecommand \bibitemStop [0]{}%
\providecommand \bibitemNoStop [0]{.\EOS\space}%
\providecommand \EOS [0]{\spacefactor3000\relax}%
\providecommand \BibitemShut  [1]{\csname bibitem#1\endcsname}%
\let\auto@bib@innerbib\@empty
\bibitem [{\citenamefont {Linder}\ and\ \citenamefont
  {Robinson}(2015)}]{Linder2015}%
  \BibitemOpen
  \bibfield  {author} {\bibinfo {author} {\bibfnamefont {J.}~\bibnamefont
  {Linder}}\ and\ \bibinfo {author} {\bibfnamefont {J.~W.~A.}\ \bibnamefont
  {Robinson}},\ }\bibfield  {title} {\enquote {\bibinfo {title}
  {Superconducting spintronics},}\ }\href {\doibase 10/gc4jjv} {\bibfield
  {journal} {\bibinfo  {journal} {Nature Physics}\ }\textbf {\bibinfo {volume}
  {11}},\ \bibinfo {pages} {307--315} (\bibinfo {year} {2015})}\BibitemShut
  {NoStop}%
\bibitem [{\citenamefont {Tinkham}(1996)}]{Tinkham1996}%
  \BibitemOpen
  \bibfield  {author} {\bibinfo {author} {\bibfnamefont {M.}~\bibnamefont
  {Tinkham}},\ }\href@noop {} {\emph {\bibinfo {title} {Introduction to
  Superconductivity}}},\ \bibinfo {edition} {2nd}\ ed.\ (\bibinfo  {publisher}
  {{McGraw-Hill}},\ \bibinfo {year} {1996})\BibitemShut {NoStop}%
\bibitem [{\citenamefont {{de Gennes}}(1999)}]{DeGennes1999}%
  \BibitemOpen
  \bibfield  {author} {\bibinfo {author} {\bibfnamefont {P.~G.}\ \bibnamefont
  {{de Gennes}}},\ }\href@noop {} {\emph {\bibinfo {title} {Superconductivity
  of Metals and Alloys}}},\ Advanced Book Classics\ (\bibinfo  {publisher}
  {{Westview Press}},\ \bibinfo {year} {1999})\BibitemShut {NoStop}%
\bibitem [{\citenamefont {Fossheim}\ and\ \citenamefont
  {Sudb\o}(2004)}]{Fossheim2004}%
  \BibitemOpen
  \bibfield  {author} {\bibinfo {author} {\bibfnamefont {K.}~\bibnamefont
  {Fossheim}}\ and\ \bibinfo {author} {\bibfnamefont {A.}~\bibnamefont
  {Sudb\o}},\ }\href {\doibase 10.1002/0470020784} {\emph {\bibinfo {title}
  {Superconductivity: Physics and Applications}}}\ (\bibinfo  {publisher}
  {{John Wiley \& Sons}},\ \bibinfo {year} {2004})\BibitemShut {NoStop}%
\bibitem [{\citenamefont {Eschrig}\ and\ \citenamefont
  {L\"ofwander}(2008)}]{Eschrig2008}%
  \BibitemOpen
  \bibfield  {author} {\bibinfo {author} {\bibfnamefont {M.}~\bibnamefont
  {Eschrig}}\ and\ \bibinfo {author} {\bibfnamefont {T.}~\bibnamefont
  {L\"ofwander}},\ }\bibfield  {title} {\enquote {\bibinfo {title} {Triplet
  supercurrents in clean and disordered half-metallic ferromagnets},}\ }\href
  {\doibase 10/fhgjzd} {\bibfield  {journal} {\bibinfo  {journal} {Nature
  Physics}\ }\textbf {\bibinfo {volume} {4}},\ \bibinfo {pages} {138--143}
  (\bibinfo {year} {2008})}\BibitemShut {NoStop}%
\bibitem [{\citenamefont {Eschrig}(2011)}]{Eschrig2011}%
  \BibitemOpen
  \bibfield  {author} {\bibinfo {author} {\bibfnamefont {M.}~\bibnamefont
  {Eschrig}},\ }\bibfield  {title} {\enquote {\bibinfo {title} {Spin-polarized
  supercurrents for spintronics},}\ }\href {\doibase 10/d8cj2w} {\bibfield
  {journal} {\bibinfo  {journal} {Physics Today}\ }\textbf {\bibinfo {volume}
  {64}},\ \bibinfo {pages} {43} (\bibinfo {year} {2011})}\BibitemShut {NoStop}%
\bibitem [{\citenamefont {Eschrig}(2015)}]{Eschrig2015}%
  \BibitemOpen
  \bibfield  {author} {\bibinfo {author} {\bibfnamefont {M.}~\bibnamefont
  {Eschrig}},\ }\bibfield  {title} {\enquote {\bibinfo {title} {Spin-polarized
  supercurrents for spintronics: A review of current progress},}\ }\href
  {\doibase 10.1088/0034-4885/78/10/104501} {\bibfield  {journal} {\bibinfo
  {journal} {Reports on Progress in Physics}\ }\textbf {\bibinfo {volume}
  {78}},\ \bibinfo {pages} {104501} (\bibinfo {year} {2015})}\BibitemShut
  {NoStop}%
\bibitem [{\citenamefont {Fulde}\ and\ \citenamefont
  {Ferrell}(1964)}]{Fulde1964}%
  \BibitemOpen
  \bibfield  {author} {\bibinfo {author} {\bibfnamefont {P.}~\bibnamefont
  {Fulde}}\ and\ \bibinfo {author} {\bibfnamefont {R.~A.}\ \bibnamefont
  {Ferrell}},\ }\bibfield  {title} {\enquote {\bibinfo {title}
  {Superconductivity in a strong spin-exchange field},}\ }\href {\doibase
  10/bjhc54} {\bibfield  {journal} {\bibinfo  {journal} {Physical Review}\
  }\textbf {\bibinfo {volume} {135}},\ \bibinfo {pages} {A550--A563} (\bibinfo
  {year} {1964})}\BibitemShut {NoStop}%
\bibitem [{\citenamefont {Larkin}\ and\ \citenamefont
  {Ovchinnikov}(1965)}]{Larkin1965}%
  \BibitemOpen
  \bibfield  {author} {\bibinfo {author} {\bibfnamefont {A.~I.}\ \bibnamefont
  {Larkin}}\ and\ \bibinfo {author} {\bibfnamefont {Yu.~N.}\ \bibnamefont
  {Ovchinnikov}},\ }\bibfield  {title} {\enquote {\bibinfo {title}
  {Inhomogeneous state of superconductors},}\ }\href@noop {} {\bibfield
  {journal} {\bibinfo  {journal} {Soviet Physics JETP}\ }\textbf {\bibinfo
  {volume} {20}},\ \bibinfo {pages} {762} (\bibinfo {year} {1965})}\BibitemShut
  {NoStop}%
\bibitem [{\citenamefont {Bergeret}\ \emph {et~al.}(2001)\citenamefont
  {Bergeret}, \citenamefont {Volkov},\ and\ \citenamefont
  {Efetov}}]{Bergeret2001}%
  \BibitemOpen
  \bibfield  {author} {\bibinfo {author} {\bibfnamefont {F.~S.}\ \bibnamefont
  {Bergeret}}, \bibinfo {author} {\bibfnamefont {A.~F.}\ \bibnamefont
  {Volkov}}, \ and\ \bibinfo {author} {\bibfnamefont {K.~B.}\ \bibnamefont
  {Efetov}},\ }\bibfield  {title} {\enquote {\bibinfo {title} {Long-range
  proximity effects in superconductor-ferromagnet structures},}\ }\href
  {\doibase 10/bg59wr} {\bibfield  {journal} {\bibinfo  {journal} {Physical
  Review Letters}\ }\textbf {\bibinfo {volume} {86}},\ \bibinfo {pages}
  {4096--4099} (\bibinfo {year} {2001})}\BibitemShut {NoStop}%
\bibitem [{\citenamefont {Bergeret}\ \emph {et~al.}(2005)\citenamefont
  {Bergeret}, \citenamefont {Volkov},\ and\ \citenamefont
  {Efetov}}]{Bergeret2005}%
  \BibitemOpen
  \bibfield  {author} {\bibinfo {author} {\bibfnamefont {F.~S.}\ \bibnamefont
  {Bergeret}}, \bibinfo {author} {\bibfnamefont {A.~F.}\ \bibnamefont
  {Volkov}}, \ and\ \bibinfo {author} {\bibfnamefont {K.~B.}\ \bibnamefont
  {Efetov}},\ }\bibfield  {title} {\enquote {\bibinfo {title} {Odd triplet
  superconductivity and related phenomena in superconductor-ferromagnet
  structures},}\ }\href {\doibase 10/bqc4vn} {\bibfield  {journal} {\bibinfo
  {journal} {Reviews of Modern Physics}\ }\textbf {\bibinfo {volume} {77}},\
  \bibinfo {pages} {1321--1373} (\bibinfo {year} {2005})}\BibitemShut {NoStop}%
\bibitem [{\citenamefont {Buzdin}(2005)}]{Buzdin2005}%
  \BibitemOpen
  \bibfield  {author} {\bibinfo {author} {\bibfnamefont {A.~I.}\ \bibnamefont
  {Buzdin}},\ }\bibfield  {title} {\enquote {\bibinfo {title} {Proximity
  effects in superconductor-ferromagnet heterostructures},}\ }\href {\doibase
  10/c3w2rr} {\bibfield  {journal} {\bibinfo  {journal} {Reviews of Modern
  Physics}\ }\textbf {\bibinfo {volume} {77}},\ \bibinfo {pages} {935--976}
  (\bibinfo {year} {2005})}\BibitemShut {NoStop}%
\bibitem [{\citenamefont {Bergeret}\ and\ \citenamefont
  {Tokatly}(2013)}]{Bergeret2013}%
  \BibitemOpen
  \bibfield  {author} {\bibinfo {author} {\bibfnamefont {F.~S.}\ \bibnamefont
  {Bergeret}}\ and\ \bibinfo {author} {\bibfnamefont {I.~V.}\ \bibnamefont
  {Tokatly}},\ }\bibfield  {title} {\enquote {\bibinfo {title} {Singlet-triplet
  conversion and the long-range proximity effect in superconductor-ferromagnet
  structures with generic spin dependent fields},}\ }\href {\doibase 10/gc5phk}
  {\bibfield  {journal} {\bibinfo  {journal} {Physical Review Letters}\
  }\textbf {\bibinfo {volume} {110}},\ \bibinfo {pages} {117003} (\bibinfo
  {year} {2013})}\BibitemShut {NoStop}%
\bibitem [{\citenamefont {Bergeret}\ and\ \citenamefont
  {Tokatly}(2014)}]{Bergeret2014}%
  \BibitemOpen
  \bibfield  {author} {\bibinfo {author} {\bibfnamefont {F.~S.}\ \bibnamefont
  {Bergeret}}\ and\ \bibinfo {author} {\bibfnamefont {I.~V.}\ \bibnamefont
  {Tokatly}},\ }\bibfield  {title} {\enquote {\bibinfo {title} {Spin-orbit
  coupling as a source of long-range triplet proximity effect in
  superconductor-ferromagnet hybrid structures},}\ }\href {\doibase 10/gc5phg}
  {\bibfield  {journal} {\bibinfo  {journal} {Physical Review B}\ }\textbf
  {\bibinfo {volume} {89}},\ \bibinfo {pages} {134517} (\bibinfo {year}
  {2014})}\BibitemShut {NoStop}%
\bibitem [{\citenamefont {Mineev}\ and\ \citenamefont
  {Volovik}(1992)}]{Mineev1992}%
  \BibitemOpen
  \bibfield  {author} {\bibinfo {author} {\bibfnamefont {V.~P.}\ \bibnamefont
  {Mineev}}\ and\ \bibinfo {author} {\bibfnamefont {G.~E.}\ \bibnamefont
  {Volovik}},\ }\bibfield  {title} {\enquote {\bibinfo {title} {Electric dipole
  moment and spin supercurrent in superfluid {{$^3$He}}},}\ }\href {\doibase
  10/b3rnzj} {\bibfield  {journal} {\bibinfo  {journal} {Journal of Low
  Temperature Physics}\ }\textbf {\bibinfo {volume} {89}},\ \bibinfo {pages}
  {823--830} (\bibinfo {year} {1992})}\BibitemShut {NoStop}%
\bibitem [{\citenamefont {Meier}\ and\ \citenamefont {Loss}(2003)}]{Meier2003}%
  \BibitemOpen
  \bibfield  {author} {\bibinfo {author} {\bibfnamefont {F.}~\bibnamefont
  {Meier}}\ and\ \bibinfo {author} {\bibfnamefont {D.}~\bibnamefont {Loss}},\
  }\bibfield  {title} {\enquote {\bibinfo {title} {Magnetization transport and
  quantized spin conductance},}\ }\href {\doibase 10/dvvptc} {\bibfield
  {journal} {\bibinfo  {journal} {Physical Review Letters}\ }\textbf {\bibinfo
  {volume} {90}},\ \bibinfo {pages} {167204} (\bibinfo {year}
  {2003})}\BibitemShut {NoStop}%
\bibitem [{\citenamefont {Sonin}(2010)}]{Sonin2010}%
  \BibitemOpen
  \bibfield  {author} {\bibinfo {author} {\bibfnamefont {E.B.}\ \bibnamefont
  {Sonin}},\ }\bibfield  {title} {\enquote {\bibinfo {title} {Spin currents and
  spin superfluidity},}\ }\href {\doibase 10/cjrdrr} {\bibfield  {journal}
  {\bibinfo  {journal} {Advances in Physics}\ }\textbf {\bibinfo {volume}
  {59}},\ \bibinfo {pages} {181--255} (\bibinfo {year} {2010})}\BibitemShut
  {NoStop}%
\bibitem [{\citenamefont {Sonin}(2007)}]{Sonin2007}%
  \BibitemOpen
  \bibfield  {author} {\bibinfo {author} {\bibfnamefont {E.~B.}\ \bibnamefont
  {Sonin}},\ }\bibfield  {title} {\enquote {\bibinfo {title} {Proposal for
  measuring mechanically equilibrium spin currents in the {{Rashba}} medium},}\
  }\href {\doibase 10/drxkps} {\bibfield  {journal} {\bibinfo  {journal}
  {Physical Review Letters}\ }\textbf {\bibinfo {volume} {99}},\ \bibinfo
  {pages} {266602} (\bibinfo {year} {2007})}\BibitemShut {NoStop}%
\bibitem [{\citenamefont {Kulagina}\ and\ \citenamefont
  {Linder}(2014)}]{Kulagina2014}%
  \BibitemOpen
  \bibfield  {author} {\bibinfo {author} {\bibfnamefont {I.}~\bibnamefont
  {Kulagina}}\ and\ \bibinfo {author} {\bibfnamefont {J.}~\bibnamefont
  {Linder}},\ }\bibfield  {title} {\enquote {\bibinfo {title} {Spin
  supercurrent, magnetization dynamics, and {$\varphi$}-state in spin-textured
  {{Josephson}} junctions},}\ }\href {\doibase 10/gffh3f} {\bibfield  {journal}
  {\bibinfo  {journal} {Physical Review B}\ }\textbf {\bibinfo {volume} {90}},\
  \bibinfo {pages} {054504} (\bibinfo {year} {2014})}\BibitemShut {NoStop}%
\bibitem [{\citenamefont {Bobkova}\ \emph {et~al.}(2018)\citenamefont
  {Bobkova}, \citenamefont {Bobkov},\ and\ \citenamefont
  {Silaev}}]{Bobkova2018}%
  \BibitemOpen
  \bibfield  {author} {\bibinfo {author} {\bibfnamefont {I.~V.}\ \bibnamefont
  {Bobkova}}, \bibinfo {author} {\bibfnamefont {A.~M.}\ \bibnamefont {Bobkov}},
  \ and\ \bibinfo {author} {\bibfnamefont {M.~A.}\ \bibnamefont {Silaev}},\
  }\bibfield  {title} {\enquote {\bibinfo {title} {Spin torques and magnetic
  texture dynamics driven by the supercurrent in superconductor/ferromagnet
  structures},}\ }\href {\doibase 10/gffh3h} {\bibfield  {journal} {\bibinfo
  {journal} {Physical Review B}\ }\textbf {\bibinfo {volume} {98}},\ \bibinfo
  {pages} {014521} (\bibinfo {year} {2018})}\BibitemShut {NoStop}%
\bibitem [{\citenamefont {Sinova}\ \emph {et~al.}(2015)\citenamefont {Sinova},
  \citenamefont {Valenzuela}, \citenamefont {Wunderlich}, \citenamefont
  {Back},\ and\ \citenamefont {Jungwirth}}]{Sinova2015}%
  \BibitemOpen
  \bibfield  {author} {\bibinfo {author} {\bibfnamefont {J.}~\bibnamefont
  {Sinova}}, \bibinfo {author} {\bibfnamefont {S.~O.}\ \bibnamefont
  {Valenzuela}}, \bibinfo {author} {\bibfnamefont {J.}~\bibnamefont
  {Wunderlich}}, \bibinfo {author} {\bibfnamefont {C.~H.}\ \bibnamefont
  {Back}}, \ and\ \bibinfo {author} {\bibfnamefont {T.}~\bibnamefont
  {Jungwirth}},\ }\bibfield  {title} {\enquote {\bibinfo {title} {Spin {{Hall}}
  effects},}\ }\href {\doibase 10/gcpdbg} {\bibfield  {journal} {\bibinfo
  {journal} {Reviews of Modern Physics}\ }\textbf {\bibinfo {volume} {87}},\
  \bibinfo {pages} {1213--1260} (\bibinfo {year} {2015})}\BibitemShut {NoStop}%
\bibitem [{\citenamefont {Onsager}(1931{\natexlab{a}})}]{Onsager1931}%
  \BibitemOpen
  \bibfield  {author} {\bibinfo {author} {\bibfnamefont {L.}~\bibnamefont
  {Onsager}},\ }\bibfield  {title} {\enquote {\bibinfo {title} {Reciprocal
  relations in irreversible processes. {{I}}},}\ }\href {\doibase 10/dtx2zs}
  {\bibfield  {journal} {\bibinfo  {journal} {Physical Review}\ }\textbf
  {\bibinfo {volume} {37}},\ \bibinfo {pages} {405--426} (\bibinfo {year}
  {1931}{\natexlab{a}})}\BibitemShut {NoStop}%
\bibitem [{\citenamefont {Onsager}(1931{\natexlab{b}})}]{Onsager1931a}%
  \BibitemOpen
  \bibfield  {author} {\bibinfo {author} {\bibfnamefont {L.}~\bibnamefont
  {Onsager}},\ }\bibfield  {title} {\enquote {\bibinfo {title} {Reciprocal
  relations in irreversible processes. {{II}}},}\ }\href {\doibase 10/ds5z6g}
  {\bibfield  {journal} {\bibinfo  {journal} {Physical Review}\ }\textbf
  {\bibinfo {volume} {38}},\ \bibinfo {pages} {2265--2279} (\bibinfo {year}
  {1931}{\natexlab{b}})}\BibitemShut {NoStop}%
\bibitem [{\citenamefont {{de Groot}}(1951)}]{DeGroot1951}%
  \BibitemOpen
  \bibfield  {author} {\bibinfo {author} {\bibfnamefont {S.~R.}\ \bibnamefont
  {{de Groot}}},\ }\href@noop {} {\emph {\bibinfo {title} {Thermodynamics of
  Irreversible Processes}}},\ \bibinfo {series} {Selected Topics in Modern
  Physics}, Vol.~\bibinfo {volume} {3}\ (\bibinfo  {publisher} {{North-Holland
  Publishing Company}},\ \bibinfo {year} {1951})\BibitemShut {NoStop}%
\bibitem [{\citenamefont {Saitoh}\ \emph {et~al.}(2006)\citenamefont {Saitoh},
  \citenamefont {Ueda}, \citenamefont {Miyajima},\ and\ \citenamefont
  {Tatara}}]{Saitoh2006}%
  \BibitemOpen
  \bibfield  {author} {\bibinfo {author} {\bibfnamefont {E.}~\bibnamefont
  {Saitoh}}, \bibinfo {author} {\bibfnamefont {M.}~\bibnamefont {Ueda}},
  \bibinfo {author} {\bibfnamefont {H.}~\bibnamefont {Miyajima}}, \ and\
  \bibinfo {author} {\bibfnamefont {G.}~\bibnamefont {Tatara}},\ }\bibfield
  {title} {\enquote {\bibinfo {title} {Conversion of spin current into charge
  current at room temperature: Inverse spin-{{Hall}} effect},}\ }\href
  {\doibase 10/d7x366} {\bibfield  {journal} {\bibinfo  {journal} {Applied
  Physics Letters}\ }\textbf {\bibinfo {volume} {88}},\ \bibinfo {pages}
  {182509} (\bibinfo {year} {2006})}\BibitemShut {NoStop}%
\bibitem [{\citenamefont {Uchida}\ \emph {et~al.}(2008)\citenamefont {Uchida},
  \citenamefont {Takahashi}, \citenamefont {Harii}, \citenamefont {Ieda},
  \citenamefont {Koshibae}, \citenamefont {Ando}, \citenamefont {Maekawa},\
  and\ \citenamefont {Saitoh}}]{Uchida2008}%
  \BibitemOpen
  \bibfield  {author} {\bibinfo {author} {\bibfnamefont {K.}~\bibnamefont
  {Uchida}}, \bibinfo {author} {\bibfnamefont {S.}~\bibnamefont {Takahashi}},
  \bibinfo {author} {\bibfnamefont {K.}~\bibnamefont {Harii}}, \bibinfo
  {author} {\bibfnamefont {J.}~\bibnamefont {Ieda}}, \bibinfo {author}
  {\bibfnamefont {W.}~\bibnamefont {Koshibae}}, \bibinfo {author}
  {\bibfnamefont {K.}~\bibnamefont {Ando}}, \bibinfo {author} {\bibfnamefont
  {S.}~\bibnamefont {Maekawa}}, \ and\ \bibinfo {author} {\bibfnamefont
  {E.}~\bibnamefont {Saitoh}},\ }\bibfield  {title} {\enquote {\bibinfo {title}
  {Observation of the spin {{Seebeck}} effect},}\ }\href {\doibase 10/c3mtgc}
  {\bibfield  {journal} {\bibinfo  {journal} {Nature}\ }\textbf {\bibinfo
  {volume} {455}},\ \bibinfo {pages} {778--781} (\bibinfo {year}
  {2008})}\BibitemShut {NoStop}%
\bibitem [{\citenamefont {Huang}\ \emph {et~al.}(2012)\citenamefont {Huang},
  \citenamefont {Fan}, \citenamefont {Qu}, \citenamefont {Chen}, \citenamefont
  {Wang}, \citenamefont {Wu}, \citenamefont {Chen}, \citenamefont {Xiao},\ and\
  \citenamefont {Chien}}]{Huang2012}%
  \BibitemOpen
  \bibfield  {author} {\bibinfo {author} {\bibfnamefont {S.~Y.}\ \bibnamefont
  {Huang}}, \bibinfo {author} {\bibfnamefont {X.}~\bibnamefont {Fan}}, \bibinfo
  {author} {\bibfnamefont {D.}~\bibnamefont {Qu}}, \bibinfo {author}
  {\bibfnamefont {Y.~P.}\ \bibnamefont {Chen}}, \bibinfo {author}
  {\bibfnamefont {W.~G.}\ \bibnamefont {Wang}}, \bibinfo {author}
  {\bibfnamefont {J.}~\bibnamefont {Wu}}, \bibinfo {author} {\bibfnamefont
  {T.~Y.}\ \bibnamefont {Chen}}, \bibinfo {author} {\bibfnamefont {J.~Q.}\
  \bibnamefont {Xiao}}, \ and\ \bibinfo {author} {\bibfnamefont {C.~L.}\
  \bibnamefont {Chien}},\ }\bibfield  {title} {\enquote {\bibinfo {title}
  {Transport magnetic proximity effects in platinum},}\ }\href {\doibase
  10/gd875g} {\bibfield  {journal} {\bibinfo  {journal} {Physical Review
  Letters}\ }\textbf {\bibinfo {volume} {109}},\ \bibinfo {pages} {107204}
  (\bibinfo {year} {2012})}\BibitemShut {NoStop}%
\bibitem [{\citenamefont {Weiler}\ \emph {et~al.}(2012)\citenamefont {Weiler},
  \citenamefont {Althammer}, \citenamefont {Czeschka}, \citenamefont {Huebl},
  \citenamefont {Wagner}, \citenamefont {Opel}, \citenamefont {Imort},
  \citenamefont {Reiss}, \citenamefont {Thomas}, \citenamefont {Gross},\ and\
  \citenamefont {Goennenwein}}]{Weiler2012}%
  \BibitemOpen
  \bibfield  {author} {\bibinfo {author} {\bibfnamefont {M.}~\bibnamefont
  {Weiler}}, \bibinfo {author} {\bibfnamefont {M.}~\bibnamefont {Althammer}},
  \bibinfo {author} {\bibfnamefont {Franz~D.}\ \bibnamefont {Czeschka}},
  \bibinfo {author} {\bibfnamefont {H.}~\bibnamefont {Huebl}}, \bibinfo
  {author} {\bibfnamefont {M.~S.}\ \bibnamefont {Wagner}}, \bibinfo {author}
  {\bibfnamefont {M.}~\bibnamefont {Opel}}, \bibinfo {author} {\bibfnamefont
  {I.-M.}\ \bibnamefont {Imort}}, \bibinfo {author} {\bibfnamefont
  {G.}~\bibnamefont {Reiss}}, \bibinfo {author} {\bibfnamefont
  {A.}~\bibnamefont {Thomas}}, \bibinfo {author} {\bibfnamefont
  {R.}~\bibnamefont {Gross}}, \ and\ \bibinfo {author} {\bibfnamefont
  {S.~T.~B.}\ \bibnamefont {Goennenwein}},\ }\bibfield  {title} {\enquote
  {\bibinfo {title} {Local charge and spin currents in magnetothermal
  landscapes},}\ }\href {\doibase 10/gd875h} {\bibfield  {journal} {\bibinfo
  {journal} {Physical Review Letters}\ }\textbf {\bibinfo {volume} {108}},\
  \bibinfo {pages} {106602} (\bibinfo {year} {2012})}\BibitemShut {NoStop}%
\bibitem [{\citenamefont {Ando}\ \emph {et~al.}(2008)\citenamefont {Ando},
  \citenamefont {Takahashi}, \citenamefont {Harii}, \citenamefont {Sasage},
  \citenamefont {Ieda}, \citenamefont {Maekawa},\ and\ \citenamefont
  {Saitoh}}]{Ando2008}%
  \BibitemOpen
  \bibfield  {author} {\bibinfo {author} {\bibfnamefont {K.}~\bibnamefont
  {Ando}}, \bibinfo {author} {\bibfnamefont {S.}~\bibnamefont {Takahashi}},
  \bibinfo {author} {\bibfnamefont {K.}~\bibnamefont {Harii}}, \bibinfo
  {author} {\bibfnamefont {K.}~\bibnamefont {Sasage}}, \bibinfo {author}
  {\bibfnamefont {J.}~\bibnamefont {Ieda}}, \bibinfo {author} {\bibfnamefont
  {S.}~\bibnamefont {Maekawa}}, \ and\ \bibinfo {author} {\bibfnamefont
  {E.}~\bibnamefont {Saitoh}},\ }\bibfield  {title} {\enquote {\bibinfo {title}
  {Electric manipulation of spin relaxation using the spin {{Hall}} effect},}\
  }\href {\doibase 10/dwsx4j} {\bibfield  {journal} {\bibinfo  {journal}
  {Physical Review Letters}\ }\textbf {\bibinfo {volume} {101}},\ \bibinfo
  {pages} {036601} (\bibinfo {year} {2008})}\BibitemShut {NoStop}%
\bibitem [{\citenamefont {Kontani}\ \emph {et~al.}(2009)\citenamefont
  {Kontani}, \citenamefont {Goryo},\ and\ \citenamefont
  {Hirashima}}]{Kontani2009}%
  \BibitemOpen
  \bibfield  {author} {\bibinfo {author} {\bibfnamefont {H.}~\bibnamefont
  {Kontani}}, \bibinfo {author} {\bibfnamefont {J.}~\bibnamefont {Goryo}}, \
  and\ \bibinfo {author} {\bibfnamefont {D.~S.}\ \bibnamefont {Hirashima}},\
  }\bibfield  {title} {\enquote {\bibinfo {title} {Intrinsic spin {{Hall}}
  effect in the \emph{s}-wave superconducting state: Analysis of the {{Rashba}}
  model},}\ }\href {\doibase 10/dzkw59} {\bibfield  {journal} {\bibinfo
  {journal} {Physical Review Letters}\ }\textbf {\bibinfo {volume} {102}},\
  \bibinfo {pages} {086602} (\bibinfo {year} {2009})}\BibitemShut {NoStop}%
\bibitem [{\citenamefont {Mal'shukov}\ and\ \citenamefont
  {Chu}(2011)}]{Malshukov2011}%
  \BibitemOpen
  \bibfield  {author} {\bibinfo {author} {\bibfnamefont {A.~G.}\ \bibnamefont
  {Mal'shukov}}\ and\ \bibinfo {author} {\bibfnamefont {C.~S.}\ \bibnamefont
  {Chu}},\ }\bibfield  {title} {\enquote {\bibinfo {title} {Spin-{{Hall}}
  current and spin polarization in an electrically biased {{SNS}} {{Josephson}}
  junction},}\ }\href {\doibase 10/bhgdgw} {\bibfield  {journal} {\bibinfo
  {journal} {Physical Review B}\ }\textbf {\bibinfo {volume} {84}},\ \bibinfo
  {pages} {054520} (\bibinfo {year} {2011})}\BibitemShut {NoStop}%
\bibitem [{\citenamefont {Pandey}\ \emph {et~al.}(2012)\citenamefont {Pandey},
  \citenamefont {Kontani}, \citenamefont {Hirashima}, \citenamefont {Arita},\
  and\ \citenamefont {Aoki}}]{Pandey2012}%
  \BibitemOpen
  \bibfield  {author} {\bibinfo {author} {\bibfnamefont {S.}~\bibnamefont
  {Pandey}}, \bibinfo {author} {\bibfnamefont {H.}~\bibnamefont {Kontani}},
  \bibinfo {author} {\bibfnamefont {D.~S.}\ \bibnamefont {Hirashima}}, \bibinfo
  {author} {\bibfnamefont {R.}~\bibnamefont {Arita}}, \ and\ \bibinfo {author}
  {\bibfnamefont {H.}~\bibnamefont {Aoki}},\ }\bibfield  {title} {\enquote
  {\bibinfo {title} {Spin {{Hall}} effect in iron-based superconductors: A
  {{Dirac}}-point effect},}\ }\href {\doibase 10/gd4rkw} {\bibfield  {journal}
  {\bibinfo  {journal} {Physical Review B}\ }\textbf {\bibinfo {volume} {86}},\
  \bibinfo {pages} {060507(R)} (\bibinfo {year} {2012})}\BibitemShut {NoStop}%
\bibitem [{\citenamefont {Mal'shukov}(2017)}]{Malshukov2017}%
  \BibitemOpen
  \bibfield  {author} {\bibinfo {author} {\bibfnamefont {A.~G.}\ \bibnamefont
  {Mal'shukov}},\ }\bibfield  {title} {\enquote {\bibinfo {title} {Supercurrent
  generation by spin injection in an \emph{s}-wave
  superconductor\textendash{{Rashba}} metal bilayer},}\ }\href {\doibase
  10/gd4rk5} {\bibfield  {journal} {\bibinfo  {journal} {Physical Review B}\
  }\textbf {\bibinfo {volume} {95}},\ \bibinfo {pages} {064517} (\bibinfo
  {year} {2017})}\BibitemShut {NoStop}%
\bibitem [{\citenamefont {Espedal}\ \emph {et~al.}(2017)\citenamefont
  {Espedal}, \citenamefont {Lange}, \citenamefont {Sadjina}, \citenamefont
  {Mal'shukov},\ and\ \citenamefont {Brataas}}]{Espedal2017}%
  \BibitemOpen
  \bibfield  {author} {\bibinfo {author} {\bibfnamefont {C.}~\bibnamefont
  {Espedal}}, \bibinfo {author} {\bibfnamefont {P.}~\bibnamefont {Lange}},
  \bibinfo {author} {\bibfnamefont {S.}~\bibnamefont {Sadjina}}, \bibinfo
  {author} {\bibfnamefont {A.~G.}\ \bibnamefont {Mal'shukov}}, \ and\ \bibinfo
  {author} {\bibfnamefont {A.}~\bibnamefont {Brataas}},\ }\bibfield  {title}
  {\enquote {\bibinfo {title} {Spin {{Hall}} effect and spin swapping in
  diffusive superconductors},}\ }\href {\doibase 10/gd4vdj} {\bibfield
  {journal} {\bibinfo  {journal} {Physical Review B}\ }\textbf {\bibinfo
  {volume} {95}},\ \bibinfo {pages} {054509} (\bibinfo {year}
  {2017})}\BibitemShut {NoStop}%
\bibitem [{\citenamefont {Wakamura}\ \emph {et~al.}(2015)\citenamefont
  {Wakamura}, \citenamefont {Akaike}, \citenamefont {Omori}, \citenamefont
  {Niimi}, \citenamefont {Takahashi}, \citenamefont {Fujimaki}, \citenamefont
  {Maekawa},\ and\ \citenamefont {Otani}}]{Wakamura2015}%
  \BibitemOpen
  \bibfield  {author} {\bibinfo {author} {\bibfnamefont {T.}~\bibnamefont
  {Wakamura}}, \bibinfo {author} {\bibfnamefont {H.}~\bibnamefont {Akaike}},
  \bibinfo {author} {\bibfnamefont {Y.}~\bibnamefont {Omori}}, \bibinfo
  {author} {\bibfnamefont {Y.}~\bibnamefont {Niimi}}, \bibinfo {author}
  {\bibfnamefont {S.}~\bibnamefont {Takahashi}}, \bibinfo {author}
  {\bibfnamefont {A.}~\bibnamefont {Fujimaki}}, \bibinfo {author}
  {\bibfnamefont {S.}~\bibnamefont {Maekawa}}, \ and\ \bibinfo {author}
  {\bibfnamefont {Y.}~\bibnamefont {Otani}},\ }\bibfield  {title} {\enquote
  {\bibinfo {title} {Quasiparticle-mediated spin {{Hall}} effect in a
  superconductor},}\ }\href {\doibase 10/f7gpd4} {\bibfield  {journal}
  {\bibinfo  {journal} {Nature Materials}\ }\textbf {\bibinfo {volume} {14}},\
  \bibinfo {pages} {675--678} (\bibinfo {year} {2015})}\BibitemShut {NoStop}%
\bibitem [{\citenamefont {Sengupta}\ \emph {et~al.}(2006)\citenamefont
  {Sengupta}, \citenamefont {Roy},\ and\ \citenamefont {Maiti}}]{Sengupta2006}%
  \BibitemOpen
  \bibfield  {author} {\bibinfo {author} {\bibfnamefont {K.}~\bibnamefont
  {Sengupta}}, \bibinfo {author} {\bibfnamefont {R.}~\bibnamefont {Roy}}, \
  and\ \bibinfo {author} {\bibfnamefont {M.}~\bibnamefont {Maiti}},\ }\bibfield
   {title} {\enquote {\bibinfo {title} {Spin {{Hall}} effect in triplet chiral
  superconductors and graphene},}\ }\href {\doibase 10/bxt977} {\bibfield
  {journal} {\bibinfo  {journal} {Physical Review B}\ }\textbf {\bibinfo
  {volume} {74}},\ \bibinfo {pages} {094505} (\bibinfo {year}
  {2006})}\BibitemShut {NoStop}%
\bibitem [{\citenamefont {Mal'shukov}\ \emph {et~al.}(2010)\citenamefont
  {Mal'shukov}, \citenamefont {Sadjina},\ and\ \citenamefont
  {Brataas}}]{Malshukov2010}%
  \BibitemOpen
  \bibfield  {author} {\bibinfo {author} {\bibfnamefont {A.~G.}\ \bibnamefont
  {Mal'shukov}}, \bibinfo {author} {\bibfnamefont {S.}~\bibnamefont {Sadjina}},
  \ and\ \bibinfo {author} {\bibfnamefont {A.}~\bibnamefont {Brataas}},\
  }\bibfield  {title} {\enquote {\bibinfo {title} {Inverse spin {{Hall}} effect
  in superconductor/normal-metal/superconductor {{Josephson}} junctions},}\
  }\href {\doibase 10/fjm2m7} {\bibfield  {journal} {\bibinfo  {journal}
  {Physical Review B}\ }\textbf {\bibinfo {volume} {81}},\ \bibinfo {pages}
  {060502(R)} (\bibinfo {year} {2010})}\BibitemShut {NoStop}%
\bibitem [{\citenamefont {Bergeret}\ and\ \citenamefont
  {Tokatly}(2016)}]{Bergeret2016}%
  \BibitemOpen
  \bibfield  {author} {\bibinfo {author} {\bibfnamefont {F.~S.}\ \bibnamefont
  {Bergeret}}\ and\ \bibinfo {author} {\bibfnamefont {I.~V.}\ \bibnamefont
  {Tokatly}},\ }\bibfield  {title} {\enquote {\bibinfo {title} {Manifestation
  of extrinsic spin hall effect in superconducting structures: Nondissipative
  magnetoelectric effects},}\ }\href {\doibase 10/gd4rk3} {\bibfield  {journal}
  {\bibinfo  {journal} {Physical Review B}\ }\textbf {\bibinfo {volume} {94}},\
  \bibinfo {pages} {180502(R)} (\bibinfo {year} {2016})}\BibitemShut {NoStop}%
\bibitem [{\citenamefont {Mal'shukov}\ and\ \citenamefont
  {Chu}(2008)}]{Malshukov2008}%
  \BibitemOpen
  \bibfield  {author} {\bibinfo {author} {\bibfnamefont {A.~G.}\ \bibnamefont
  {Mal'shukov}}\ and\ \bibinfo {author} {\bibfnamefont {C.~S.}\ \bibnamefont
  {Chu}},\ }\bibfield  {title} {\enquote {\bibinfo {title} {Spin {{Hall}}
  effect in a {{Josephson}} contact},}\ }\href {\doibase 10/dwgqd3} {\bibfield
  {journal} {\bibinfo  {journal} {Physical Review B}\ }\textbf {\bibinfo
  {volume} {78}},\ \bibinfo {pages} {104503} (\bibinfo {year}
  {2008})}\BibitemShut {NoStop}%
\bibitem [{\citenamefont {Yang}\ \emph {et~al.}(2012)\citenamefont {Yang},
  \citenamefont {Yang},\ and\ \citenamefont {Wang}}]{Yang2012}%
  \BibitemOpen
  \bibfield  {author} {\bibinfo {author} {\bibfnamefont {Z.-H.}\ \bibnamefont
  {Yang}}, \bibinfo {author} {\bibfnamefont {Y.-H.}\ \bibnamefont {Yang}}, \
  and\ \bibinfo {author} {\bibfnamefont {J.}~\bibnamefont {Wang}},\ }\bibfield
  {title} {\enquote {\bibinfo {title} {Interfacial spin {{Hall}} current in a
  {{Josephson}} junction with {{Rashba}} spin\textendash{}orbit coupling},}\
  }\href {\doibase 10/gd4rkt} {\bibfield  {journal} {\bibinfo  {journal}
  {Chinese Physics B}\ }\textbf {\bibinfo {volume} {21}},\ \bibinfo {pages}
  {057402} (\bibinfo {year} {2012})}\BibitemShut {NoStop}%
\bibitem [{\citenamefont {Linder}\ \emph {et~al.}(2017)\citenamefont {Linder},
  \citenamefont {Amundsen},\ and\ \citenamefont {Risingg\aa{}rd}}]{Linder2017}%
  \BibitemOpen
  \bibfield  {author} {\bibinfo {author} {\bibfnamefont {J.}~\bibnamefont
  {Linder}}, \bibinfo {author} {\bibfnamefont {M.}~\bibnamefont {Amundsen}}, \
  and\ \bibinfo {author} {\bibfnamefont {V.}~\bibnamefont {Risingg\aa{}rd}},\
  }\bibfield  {title} {\enquote {\bibinfo {title} {Intrinsic superspin {{Hall}}
  current},}\ }\href {\doibase 10/gc5rrm} {\bibfield  {journal} {\bibinfo
  {journal} {Physical Review B}\ }\textbf {\bibinfo {volume} {96}},\ \bibinfo
  {pages} {094512} (\bibinfo {year} {2017})}\BibitemShut {NoStop}%
\bibitem [{\citenamefont {Konschelle}\ \emph {et~al.}(2015)\citenamefont
  {Konschelle}, \citenamefont {Tokatly},\ and\ \citenamefont
  {Bergeret}}]{Konschelle2015}%
  \BibitemOpen
  \bibfield  {author} {\bibinfo {author} {\bibfnamefont {F.}~\bibnamefont
  {Konschelle}}, \bibinfo {author} {\bibfnamefont {I.~V.}\ \bibnamefont
  {Tokatly}}, \ and\ \bibinfo {author} {\bibfnamefont {F.~S.}\ \bibnamefont
  {Bergeret}},\ }\bibfield  {title} {\enquote {\bibinfo {title} {Theory of the
  spin-galvanic effect and the anomalous phase shift {$\varphi{}_{0}$} in
  superconductors and {{Josephson}} junctions with intrinsic spin-orbit
  coupling},}\ }\href {\doibase 10/gfb35s} {\bibfield  {journal} {\bibinfo
  {journal} {Physical Review B}\ }\textbf {\bibinfo {volume} {92}},\ \bibinfo
  {pages} {125443} (\bibinfo {year} {2015})}\BibitemShut {NoStop}%
\bibitem [{\citenamefont {Keizer}\ \emph {et~al.}(2006)\citenamefont {Keizer},
  \citenamefont {Goennenwein}, \citenamefont {Klapwijk}, \citenamefont {Miao},
  \citenamefont {Xiao},\ and\ \citenamefont {Gupta}}]{Keizer2006}%
  \BibitemOpen
  \bibfield  {author} {\bibinfo {author} {\bibfnamefont {R.~S.}\ \bibnamefont
  {Keizer}}, \bibinfo {author} {\bibfnamefont {S.~T.~B.}\ \bibnamefont
  {Goennenwein}}, \bibinfo {author} {\bibfnamefont {T.~M.}\ \bibnamefont
  {Klapwijk}}, \bibinfo {author} {\bibfnamefont {G.}~\bibnamefont {Miao}},
  \bibinfo {author} {\bibfnamefont {G.}~\bibnamefont {Xiao}}, \ and\ \bibinfo
  {author} {\bibfnamefont {A.}~\bibnamefont {Gupta}},\ }\bibfield  {title}
  {\enquote {\bibinfo {title} {A spin triplet supercurrent through the
  half-metallic ferromagnet {{CrO$_{2}$}}},}\ }\href {\doibase 10/bxgrgd}
  {\bibfield  {journal} {\bibinfo  {journal} {Nature}\ }\textbf {\bibinfo
  {volume} {439}},\ \bibinfo {pages} {825--827} (\bibinfo {year}
  {2006})}\BibitemShut {NoStop}%
\bibitem [{\citenamefont {Khaire}\ \emph {et~al.}(2010)\citenamefont {Khaire},
  \citenamefont {Khasawneh}, \citenamefont {Pratt},\ and\ \citenamefont
  {Birge}}]{Khaire2010}%
  \BibitemOpen
  \bibfield  {author} {\bibinfo {author} {\bibfnamefont {T.~S.}\ \bibnamefont
  {Khaire}}, \bibinfo {author} {\bibfnamefont {M.~A.}\ \bibnamefont
  {Khasawneh}}, \bibinfo {author} {\bibfnamefont {W.~P.}\ \bibnamefont
  {Pratt}}, \ and\ \bibinfo {author} {\bibfnamefont {N.~O.}\ \bibnamefont
  {Birge}},\ }\bibfield  {title} {\enquote {\bibinfo {title} {Observation of
  spin-triplet superconductivity in {{Co}}-based {{Josephson}} junctions},}\
  }\href {\doibase 10/bkvp93} {\bibfield  {journal} {\bibinfo  {journal}
  {Physical Review Letters}\ }\textbf {\bibinfo {volume} {104}},\ \bibinfo
  {pages} {137002} (\bibinfo {year} {2010})}\BibitemShut {NoStop}%
\bibitem [{\citenamefont {Robinson}\ \emph {et~al.}(2010)\citenamefont
  {Robinson}, \citenamefont {Witt},\ and\ \citenamefont
  {Blamire}}]{Robinson2010}%
  \BibitemOpen
  \bibfield  {author} {\bibinfo {author} {\bibfnamefont {J.~W.~A.}\
  \bibnamefont {Robinson}}, \bibinfo {author} {\bibfnamefont {J.~D.~S.}\
  \bibnamefont {Witt}}, \ and\ \bibinfo {author} {\bibfnamefont {M.~G.}\
  \bibnamefont {Blamire}},\ }\bibfield  {title} {\enquote {\bibinfo {title}
  {Controlled injection of spin-triplet supercurrents into a strong
  ferromagnet},}\ }\href {\doibase 10/d9rkh4} {\bibfield  {journal} {\bibinfo
  {journal} {Science}\ }\textbf {\bibinfo {volume} {329}},\ \bibinfo {pages}
  {59--61} (\bibinfo {year} {2010})}\BibitemShut {NoStop}%
\bibitem [{\citenamefont {Anwar}\ \emph {et~al.}(2012)\citenamefont {Anwar},
  \citenamefont {Veldhorst}, \citenamefont {Brinkman},\ and\ \citenamefont
  {Aarts}}]{Anwar2012}%
  \BibitemOpen
  \bibfield  {author} {\bibinfo {author} {\bibfnamefont {M.~S.}\ \bibnamefont
  {Anwar}}, \bibinfo {author} {\bibfnamefont {M.}~\bibnamefont {Veldhorst}},
  \bibinfo {author} {\bibfnamefont {A.}~\bibnamefont {Brinkman}}, \ and\
  \bibinfo {author} {\bibfnamefont {J.}~\bibnamefont {Aarts}},\ }\bibfield
  {title} {\enquote {\bibinfo {title} {Long range supercurrents in
  ferromagnetic {{CrO$_{2}$}} using a multilayer contact structure},}\ }\href
  {\doibase 10/gft6rn} {\bibfield  {journal} {\bibinfo  {journal} {Applied
  Physics Letters}\ }\textbf {\bibinfo {volume} {100}},\ \bibinfo {pages}
  {052602} (\bibinfo {year} {2012})}\BibitemShut {NoStop}%
\bibitem [{\citenamefont {Martinez}\ \emph {et~al.}(2016)\citenamefont
  {Martinez}, \citenamefont {Pratt},\ and\ \citenamefont
  {Birge}}]{Martinez2016}%
  \BibitemOpen
  \bibfield  {author} {\bibinfo {author} {\bibfnamefont {W.~M.}\ \bibnamefont
  {Martinez}}, \bibinfo {author} {\bibfnamefont {W.~P.}\ \bibnamefont {Pratt}},
  \ and\ \bibinfo {author} {\bibfnamefont {N.~O.}\ \bibnamefont {Birge}},\
  }\bibfield  {title} {\enquote {\bibinfo {title} {Amplitude control of the
  spin-triplet supercurrent in {{S/F/S}} {{Josephson}} junctions},}\ }\href
  {\doibase 10/gft6rp} {\bibfield  {journal} {\bibinfo  {journal} {Physical
  Review Letters}\ }\textbf {\bibinfo {volume} {116}},\ \bibinfo {pages}
  {077001} (\bibinfo {year} {2016})}\BibitemShut {NoStop}%
\bibitem [{\citenamefont {Singh}\ \emph {et~al.}(2016)\citenamefont {Singh},
  \citenamefont {Jansen}, \citenamefont {Lahabi},\ and\ \citenamefont
  {Aarts}}]{Singh2016}%
  \BibitemOpen
  \bibfield  {author} {\bibinfo {author} {\bibfnamefont {A.}~\bibnamefont
  {Singh}}, \bibinfo {author} {\bibfnamefont {C.}~\bibnamefont {Jansen}},
  \bibinfo {author} {\bibfnamefont {K.}~\bibnamefont {Lahabi}}, \ and\ \bibinfo
  {author} {\bibfnamefont {J.}~\bibnamefont {Aarts}},\ }\bibfield  {title}
  {\enquote {\bibinfo {title} {High-quality {{CrO$_{2}$}} nanowires for
  dissipation-less spintronics},}\ }\href {\doibase 10/f88q58} {\bibfield
  {journal} {\bibinfo  {journal} {Physical Review X}\ }\textbf {\bibinfo
  {volume} {6}},\ \bibinfo {pages} {041012} (\bibinfo {year}
  {2016})}\BibitemShut {NoStop}%
\bibitem [{\citenamefont {Slonczewski}(1989)}]{Slonczewski1989}%
  \BibitemOpen
  \bibfield  {author} {\bibinfo {author} {\bibfnamefont {J.~C.}\ \bibnamefont
  {Slonczewski}},\ }\bibfield  {title} {\enquote {\bibinfo {title} {Conductance
  and exchange coupling of two ferromagnets separated by a tunneling
  barrier},}\ }\href {\doibase 10/dwbbn2} {\bibfield  {journal} {\bibinfo
  {journal} {Physical Review B}\ }\textbf {\bibinfo {volume} {39}},\ \bibinfo
  {pages} {6995--7002} (\bibinfo {year} {1989})}\BibitemShut {NoStop}%
\bibitem [{\citenamefont {Chen}\ \emph {et~al.}(2014)\citenamefont {Chen},
  \citenamefont {Horsch},\ and\ \citenamefont {Manske}}]{Chen2014}%
  \BibitemOpen
  \bibfield  {author} {\bibinfo {author} {\bibfnamefont {W.}~\bibnamefont
  {Chen}}, \bibinfo {author} {\bibfnamefont {P.}~\bibnamefont {Horsch}}, \ and\
  \bibinfo {author} {\bibfnamefont {D.}~\bibnamefont {Manske}},\ }\bibfield
  {title} {\enquote {\bibinfo {title} {Dissipationless spin current between two
  coupled ferromagnets},}\ }\href {\doibase 10/gd876t} {\bibfield  {journal}
  {\bibinfo  {journal} {Physical Review B}\ }\textbf {\bibinfo {volume} {89}},\
  \bibinfo {pages} {064427} (\bibinfo {year} {2014})}\BibitemShut {NoStop}%
\bibitem [{\citenamefont {Tanaka}\ \emph {et~al.}(2007)\citenamefont {Tanaka},
  \citenamefont {Golubov}, \citenamefont {Kashiwaya},\ and\ \citenamefont
  {Ueda}}]{Tanaka2007}%
  \BibitemOpen
  \bibfield  {author} {\bibinfo {author} {\bibfnamefont {Y.}~\bibnamefont
  {Tanaka}}, \bibinfo {author} {\bibfnamefont {A.~A.}\ \bibnamefont {Golubov}},
  \bibinfo {author} {\bibfnamefont {S.}~\bibnamefont {Kashiwaya}}, \ and\
  \bibinfo {author} {\bibfnamefont {M.}~\bibnamefont {Ueda}},\ }\bibfield
  {title} {\enquote {\bibinfo {title} {Anomalous {{Josephson}} effect between
  even- and odd-frequency superconductors},}\ }\href {\doibase 10/c438x3}
  {\bibfield  {journal} {\bibinfo  {journal} {Physical Review Letters}\
  }\textbf {\bibinfo {volume} {99}},\ \bibinfo {pages} {037005} (\bibinfo
  {year} {2007})}\BibitemShut {NoStop}%
\bibitem [{\citenamefont {Eschrig}\ \emph {et~al.}(2007)\citenamefont
  {Eschrig}, \citenamefont {L\"ofwander}, \citenamefont {Champel},
  \citenamefont {Cuevas}, \citenamefont {Kopu},\ and\ \citenamefont
  {Sch\"on}}]{Eschrig2007}%
  \BibitemOpen
  \bibfield  {author} {\bibinfo {author} {\bibfnamefont {M.}~\bibnamefont
  {Eschrig}}, \bibinfo {author} {\bibfnamefont {T.}~\bibnamefont
  {L\"ofwander}}, \bibinfo {author} {\bibfnamefont {T.}~\bibnamefont
  {Champel}}, \bibinfo {author} {\bibfnamefont {J.~C.}\ \bibnamefont {Cuevas}},
  \bibinfo {author} {\bibfnamefont {J.}~\bibnamefont {Kopu}}, \ and\ \bibinfo
  {author} {\bibfnamefont {G.}~\bibnamefont {Sch\"on}},\ }\bibfield  {title}
  {\enquote {\bibinfo {title} {Symmetries of pairing correlations in
  superconductor\textendash{}ferromagnet nanostructures},}\ }\href {\doibase
  10/dpgqq3} {\bibfield  {journal} {\bibinfo  {journal} {Journal of Low
  Temperature Physics}\ }\textbf {\bibinfo {volume} {147}},\ \bibinfo {pages}
  {457--476} (\bibinfo {year} {2007})}\BibitemShut {NoStop}%
\bibitem [{\citenamefont {Sigrist}\ and\ \citenamefont
  {Ueda}(1991)}]{Sigrist1991}%
  \BibitemOpen
  \bibfield  {author} {\bibinfo {author} {\bibfnamefont {M.}~\bibnamefont
  {Sigrist}}\ and\ \bibinfo {author} {\bibfnamefont {K.}~\bibnamefont {Ueda}},\
  }\bibfield  {title} {\enquote {\bibinfo {title} {Phenomenological theory of
  unconventional superconductivity},}\ }\href {\doibase 10/fvp2gm} {\bibfield
  {journal} {\bibinfo  {journal} {Reviews of Modern Physics}\ }\textbf
  {\bibinfo {volume} {63}},\ \bibinfo {pages} {239--311} (\bibinfo {year}
  {1991})}\BibitemShut {NoStop}%
\bibitem [{\citenamefont {Dummit}\ and\ \citenamefont
  {Foote}(2004)}]{Dummit2004}%
  \BibitemOpen
  \bibfield  {author} {\bibinfo {author} {\bibfnamefont {D.~S.}\ \bibnamefont
  {Dummit}}\ and\ \bibinfo {author} {\bibfnamefont {R.~M.}\ \bibnamefont
  {Foote}},\ }\href@noop {} {\emph {\bibinfo {title} {Abstract Algebra}}},\
  \bibinfo {edition} {3rd}\ ed.\ (\bibinfo  {publisher} {{John Wiley \&
  Sons}},\ \bibinfo {year} {2004})\BibitemShut {NoStop}%
\bibitem [{\citenamefont {Gantmacher}(2000)}]{Gantmacher2000}%
  \BibitemOpen
  \bibfield  {author} {\bibinfo {author} {\bibfnamefont {F.~R.}\ \bibnamefont
  {Gantmacher}},\ }\href@noop {} {\emph {\bibinfo {title} {The Theory of
  Matrices}}},\ Vol.~\bibinfo {volume} {1}\ (\bibinfo  {publisher} {{American
  Mathematical Society}},\ \bibinfo {year} {2000})\BibitemShut {NoStop}%
\bibitem [{\citenamefont {Ouassou}\ \emph {et~al.}(2017)\citenamefont
  {Ouassou}, \citenamefont {Jacobsen},\ and\ \citenamefont
  {Linder}}]{Ouassou2017}%
  \BibitemOpen
  \bibfield  {author} {\bibinfo {author} {\bibfnamefont {J.~A.}\ \bibnamefont
  {Ouassou}}, \bibinfo {author} {\bibfnamefont {S.~H.}\ \bibnamefont
  {Jacobsen}}, \ and\ \bibinfo {author} {\bibfnamefont {J.}~\bibnamefont
  {Linder}},\ }\bibfield  {title} {\enquote {\bibinfo {title} {Conservation of
  spin supercurrents in superconductors},}\ }\href {\doibase 10/gc4m2n}
  {\bibfield  {journal} {\bibinfo  {journal} {Physical Review B}\ }\textbf
  {\bibinfo {volume} {96}},\ \bibinfo {pages} {094505} (\bibinfo {year}
  {2017})}\BibitemShut {NoStop}%
\bibitem [{\citenamefont {Leggett}(1975)}]{Leggett1975}%
  \BibitemOpen
  \bibfield  {author} {\bibinfo {author} {\bibfnamefont {A.~J.}\ \bibnamefont
  {Leggett}},\ }\bibfield  {title} {\enquote {\bibinfo {title} {A theoretical
  description of the new phases of liquid {{$^3$He}}},}\ }\href {\doibase
  10/dxsc69} {\bibfield  {journal} {\bibinfo  {journal} {Reviews of Modern
  Physics}\ }\textbf {\bibinfo {volume} {47}},\ \bibinfo {pages} {331--414}
  (\bibinfo {year} {1975})}\BibitemShut {NoStop}%
\bibitem [{\citenamefont {Coey}(2010)}]{Coey2010}%
  \BibitemOpen
  \bibfield  {author} {\bibinfo {author} {\bibfnamefont {J.~M.~D.}\
  \bibnamefont {Coey}},\ }\href {\doibase 10.1017/CBO9780511845000} {\emph
  {\bibinfo {title} {Magnetism and Magnetic Materials}}}\ (\bibinfo
  {publisher} {{Cambridge University Press}},\ \bibinfo {year}
  {2010})\BibitemShut {NoStop}%
\bibitem [{\citenamefont {O'Handley}(2000)}]{OHandley2000}%
  \BibitemOpen
  \bibfield  {author} {\bibinfo {author} {\bibfnamefont {R.~C.}\ \bibnamefont
  {O'Handley}},\ }\href@noop {} {\emph {\bibinfo {title} {Modern Magnetic
  Materials: Principles and Applications}}}\ (\bibinfo  {publisher} {{John
  Wiley \& Sons}},\ \bibinfo {year} {2000})\BibitemShut {NoStop}%
\bibitem [{\citenamefont {Liu}\ and\ \citenamefont
  {Chan}(2010{\natexlab{a}})}]{Liu2010}%
  \BibitemOpen
  \bibfield  {author} {\bibinfo {author} {\bibfnamefont {J.-F.}\ \bibnamefont
  {Liu}}\ and\ \bibinfo {author} {\bibfnamefont {K.~S.}\ \bibnamefont {Chan}},\
  }\bibfield  {title} {\enquote {\bibinfo {title} {Relation between symmetry
  breaking and the anomalous {{Josephson}} effect},}\ }\href {\doibase
  10/d5bvht} {\bibfield  {journal} {\bibinfo  {journal} {Physical Review B}\
  }\textbf {\bibinfo {volume} {82}},\ \bibinfo {pages} {125305} (\bibinfo
  {year} {2010}{\natexlab{a}})}\BibitemShut {NoStop}%
\bibitem [{\citenamefont {Liu}\ and\ \citenamefont
  {Chan}(2010{\natexlab{b}})}]{Liu2010a}%
  \BibitemOpen
  \bibfield  {author} {\bibinfo {author} {\bibfnamefont {J.-F.}\ \bibnamefont
  {Liu}}\ and\ \bibinfo {author} {\bibfnamefont {K.~S.}\ \bibnamefont {Chan}},\
  }\bibfield  {title} {\enquote {\bibinfo {title} {Anomalous {{Josephson}}
  current through a ferromagnetic trilayer junction},}\ }\href {\doibase
  10/cgnw3h} {\bibfield  {journal} {\bibinfo  {journal} {Physical Review B}\
  }\textbf {\bibinfo {volume} {82}},\ \bibinfo {pages} {184533} (\bibinfo
  {year} {2010}{\natexlab{b}})}\BibitemShut {NoStop}%
\bibitem [{\citenamefont {Szombati}\ \emph {et~al.}(2016)\citenamefont
  {Szombati}, \citenamefont {{Nadj-Perge}}, \citenamefont {Car}, \citenamefont
  {Plissard}, \citenamefont {Bakkers},\ and\ \citenamefont
  {Kouwenhoven}}]{Szombati2016}%
  \BibitemOpen
  \bibfield  {author} {\bibinfo {author} {\bibfnamefont {D.~B.}\ \bibnamefont
  {Szombati}}, \bibinfo {author} {\bibfnamefont {S.}~\bibnamefont
  {{Nadj-Perge}}}, \bibinfo {author} {\bibfnamefont {D.}~\bibnamefont {Car}},
  \bibinfo {author} {\bibfnamefont {S.~R.}\ \bibnamefont {Plissard}}, \bibinfo
  {author} {\bibfnamefont {E.~P. A.~M.}\ \bibnamefont {Bakkers}}, \ and\
  \bibinfo {author} {\bibfnamefont {L.~P.}\ \bibnamefont {Kouwenhoven}},\
  }\bibfield  {title} {\enquote {\bibinfo {title} {Josephson
  {$\varphi{}_{0}$}-junction in nanowire quantum dots},}\ }\href {\doibase
  10/f8p7r7} {\bibfield  {journal} {\bibinfo  {journal} {Nature Physics}\
  }\textbf {\bibinfo {volume} {12}},\ \bibinfo {pages} {568--572} (\bibinfo
  {year} {2016})}\BibitemShut {NoStop}%
\bibitem [{\citenamefont {Rasmussen}\ \emph {et~al.}(2016)\citenamefont
  {Rasmussen}, \citenamefont {Danon}, \citenamefont {Suominen}, \citenamefont
  {Nichele}, \citenamefont {Kjaergaard},\ and\ \citenamefont
  {Flensberg}}]{Rasmussen2016}%
  \BibitemOpen
  \bibfield  {author} {\bibinfo {author} {\bibfnamefont {A.}~\bibnamefont
  {Rasmussen}}, \bibinfo {author} {\bibfnamefont {J.}~\bibnamefont {Danon}},
  \bibinfo {author} {\bibfnamefont {H.}~\bibnamefont {Suominen}}, \bibinfo
  {author} {\bibfnamefont {F.}~\bibnamefont {Nichele}}, \bibinfo {author}
  {\bibfnamefont {M.}~\bibnamefont {Kjaergaard}}, \ and\ \bibinfo {author}
  {\bibfnamefont {K.}~\bibnamefont {Flensberg}},\ }\bibfield  {title} {\enquote
  {\bibinfo {title} {Effects of spin-orbit coupling and spatial symmetries on
  the {{Josephson}} current in {{SNS}} junctions},}\ }\href {\doibase
  10/gffjdb} {\bibfield  {journal} {\bibinfo  {journal} {Physical Review B}\
  }\textbf {\bibinfo {volume} {93}},\ \bibinfo {pages} {155406} (\bibinfo
  {year} {2016})}\BibitemShut {NoStop}%
\bibitem [{\citenamefont {Chandrasekhar}(2008)}]{Chandrasekhar2008}%
  \BibitemOpen
  \bibfield  {author} {\bibinfo {author} {\bibfnamefont {V.}~\bibnamefont
  {Chandrasekhar}},\ }\bibfield  {title} {\enquote {\bibinfo {title}
  {Proximity-coupled systems: quasiclassical theory of superconductivity},}\
  }in\ \href {\doibase 10.1007/978-3-540-73253-2_8} {\emph {\bibinfo
  {booktitle} {Superconductivity}}},\ \bibinfo {editor} {edited by\ \bibinfo
  {editor} {\bibfnamefont {K.~H.}\ \bibnamefont {Bennemann}}\ and\ \bibinfo
  {editor} {\bibfnamefont {J.~B.}\ \bibnamefont {Ketterson}}}\ (\bibinfo
  {publisher} {{Springer}},\ \bibinfo {year} {2008})\ pp.\ \bibinfo {pages}
  {279--313}\BibitemShut {NoStop}%
\bibitem [{\citenamefont {{Black-Schaffer}}\ and\ \citenamefont
  {Linder}(2010)}]{Black-Schaffer2010}%
  \BibitemOpen
  \bibfield  {author} {\bibinfo {author} {\bibfnamefont {A.~M.}\ \bibnamefont
  {{Black-Schaffer}}}\ and\ \bibinfo {author} {\bibfnamefont {J.}~\bibnamefont
  {Linder}},\ }\bibfield  {title} {\enquote {\bibinfo {title} {Strongly
  anharmonic current-phase relation in ballistic graphene {{Josephson}}
  junctions},}\ }\href {\doibase 10/cw6jqx} {\bibfield  {journal} {\bibinfo
  {journal} {Physical Review B}\ }\textbf {\bibinfo {volume} {82}},\ \bibinfo
  {pages} {184522} (\bibinfo {year} {2010})}\BibitemShut {NoStop}%
\bibitem [{\citenamefont {English}\ \emph {et~al.}(2016)\citenamefont
  {English}, \citenamefont {Hamilton}, \citenamefont {Chialvo}, \citenamefont
  {Moraru}, \citenamefont {Mason},\ and\ \citenamefont
  {Van~Harlingen}}]{English2016}%
  \BibitemOpen
  \bibfield  {author} {\bibinfo {author} {\bibfnamefont {C.~D.}\ \bibnamefont
  {English}}, \bibinfo {author} {\bibfnamefont {D.~R.}\ \bibnamefont
  {Hamilton}}, \bibinfo {author} {\bibfnamefont {C.}~\bibnamefont {Chialvo}},
  \bibinfo {author} {\bibfnamefont {I.~C.}\ \bibnamefont {Moraru}}, \bibinfo
  {author} {\bibfnamefont {N.}~\bibnamefont {Mason}}, \ and\ \bibinfo {author}
  {\bibfnamefont {D.~J.}\ \bibnamefont {Van~Harlingen}},\ }\bibfield  {title}
  {\enquote {\bibinfo {title} {Observation of nonsinusoidal current-phase
  relation in graphene {{Josephson}} junctions},}\ }\href {\doibase 10/gfvgq5}
  {\bibfield  {journal} {\bibinfo  {journal} {Physical Review B}\ }\textbf
  {\bibinfo {volume} {94}},\ \bibinfo {pages} {115435} (\bibinfo {year}
  {2016})}\BibitemShut {NoStop}%
\bibitem [{\citenamefont {Kato}\ \emph {et~al.}(2004)\citenamefont {Kato},
  \citenamefont {Myers}, \citenamefont {Gossard},\ and\ \citenamefont
  {Awschalom}}]{Kato2004}%
  \BibitemOpen
  \bibfield  {author} {\bibinfo {author} {\bibfnamefont {Y.~K.}\ \bibnamefont
  {Kato}}, \bibinfo {author} {\bibfnamefont {R.~C.}\ \bibnamefont {Myers}},
  \bibinfo {author} {\bibfnamefont {A.~C.}\ \bibnamefont {Gossard}}, \ and\
  \bibinfo {author} {\bibfnamefont {D.~D.}\ \bibnamefont {Awschalom}},\
  }\bibfield  {title} {\enquote {\bibinfo {title} {Observation of the spin
  {{Hall}} effect in semiconductors},}\ }\href {\doibase 10/fgsr2g} {\bibfield
  {journal} {\bibinfo  {journal} {Science}\ }\textbf {\bibinfo {volume}
  {306}},\ \bibinfo {pages} {1910--1913} (\bibinfo {year} {2004})}\BibitemShut
  {NoStop}%
\bibitem [{\citenamefont {Wunderlich}\ \emph {et~al.}(2005)\citenamefont
  {Wunderlich}, \citenamefont {Kaestner}, \citenamefont {Sinova},\ and\
  \citenamefont {Jungwirth}}]{Wunderlich2005}%
  \BibitemOpen
  \bibfield  {author} {\bibinfo {author} {\bibfnamefont {J.}~\bibnamefont
  {Wunderlich}}, \bibinfo {author} {\bibfnamefont {B.}~\bibnamefont
  {Kaestner}}, \bibinfo {author} {\bibfnamefont {J.}~\bibnamefont {Sinova}}, \
  and\ \bibinfo {author} {\bibfnamefont {T.}~\bibnamefont {Jungwirth}},\
  }\bibfield  {title} {\enquote {\bibinfo {title} {Experimental observation of
  the spin-{{Hall}} effect in a two-dimensional spin-orbit coupled
  semiconductor system},}\ }\href {\doibase 10/fhvg8q} {\bibfield  {journal}
  {\bibinfo  {journal} {Physical Review Letters}\ }\textbf {\bibinfo {volume}
  {94}},\ \bibinfo {pages} {047204} (\bibinfo {year} {2005})}\BibitemShut
  {NoStop}%
\bibitem [{\citenamefont {Halterman}\ \emph {et~al.}(2008)\citenamefont
  {Halterman}, \citenamefont {Valls},\ and\ \citenamefont
  {Barsic}}]{Halterman2008}%
  \BibitemOpen
  \bibfield  {author} {\bibinfo {author} {\bibfnamefont {K.}~\bibnamefont
  {Halterman}}, \bibinfo {author} {\bibfnamefont {O.~T.}\ \bibnamefont
  {Valls}}, \ and\ \bibinfo {author} {\bibfnamefont {P.~H.}\ \bibnamefont
  {Barsic}},\ }\bibfield  {title} {\enquote {\bibinfo {title} {Induced triplet
  pairing in clean \emph{s}-wave superconductor/ferromagnet layered
  structures},}\ }\href {\doibase 10/cbjpds} {\bibfield  {journal} {\bibinfo
  {journal} {Physical Review B}\ }\textbf {\bibinfo {volume} {77}},\ \bibinfo
  {pages} {174511} (\bibinfo {year} {2008})}\BibitemShut {NoStop}%
\bibitem [{\citenamefont {Terrade}\ \emph {et~al.}(2016)\citenamefont
  {Terrade}, \citenamefont {Manske},\ and\ \citenamefont
  {Cuoco}}]{Terrade2016}%
  \BibitemOpen
  \bibfield  {author} {\bibinfo {author} {\bibfnamefont {D.}~\bibnamefont
  {Terrade}}, \bibinfo {author} {\bibfnamefont {D.}~\bibnamefont {Manske}}, \
  and\ \bibinfo {author} {\bibfnamefont {M.}~\bibnamefont {Cuoco}},\ }\bibfield
   {title} {\enquote {\bibinfo {title} {Control of edge currents at a
  ferromagnet\textendash{}triplet superconductor interface by multiple helical
  modes},}\ }\href {\doibase 10/gd876b} {\bibfield  {journal} {\bibinfo
  {journal} {Physical Review B}\ }\textbf {\bibinfo {volume} {93}},\ \bibinfo
  {pages} {104523} (\bibinfo {year} {2016})}\BibitemShut {NoStop}%
\bibitem [{\citenamefont {Hikino}(2018)}]{Hikino2018}%
  \BibitemOpen
  \bibfield  {author} {\bibinfo {author} {\bibfnamefont {Shin-ichi}\
  \bibnamefont {Hikino}},\ }\bibfield  {title} {\enquote {\bibinfo {title}
  {Magnetization reversal by tuning {{Rashba}} spin\textendash{}orbit
  interaction and {{Josephson}} phase in a ferromagnetic {{Josephson}}
  junction},}\ }\href {\doibase 10/gdvkdd} {\bibfield  {journal} {\bibinfo
  {journal} {Journal of the Physical Society of Japan}\ }\textbf {\bibinfo
  {volume} {87}},\ \bibinfo {pages} {074707} (\bibinfo {year}
  {2018})}\BibitemShut {NoStop}%
\bibitem [{\citenamefont {Andreev}(1964)}]{Andreev1964}%
  \BibitemOpen
  \bibfield  {author} {\bibinfo {author} {\bibfnamefont {A.~F.}\ \bibnamefont
  {Andreev}},\ }\bibfield  {title} {\enquote {\bibinfo {title} {The thermal
  conductivity of the intermediate state in superconductors},}\ }\href@noop {}
  {\bibfield  {journal} {\bibinfo  {journal} {Soviet Physics JETP}\ }\textbf
  {\bibinfo {volume} {19}},\ \bibinfo {pages} {1228} (\bibinfo {year}
  {1964})}\BibitemShut {NoStop}%
\bibitem [{\citenamefont {Sauls}(2018)}]{Sauls2018}%
  \BibitemOpen
  \bibfield  {author} {\bibinfo {author} {\bibfnamefont {J.~A.}\ \bibnamefont
  {Sauls}},\ }\bibfield  {title} {\enquote {\bibinfo {title} {Andreev bound
  states and their signatures},}\ }\href {\doibase 10/gd876f} {\bibfield
  {journal} {\bibinfo  {journal} {Philosophical Transactions of the Royal
  Society A: Mathematical, Physical and Engineering Sciences}\ }\textbf
  {\bibinfo {volume} {376}},\ \bibinfo {pages} {20180140} (\bibinfo {year}
  {2018})}\BibitemShut {NoStop}%
\bibitem [{\citenamefont {Kulik}\ and\ \citenamefont
  {Omelyanchuk}(1977)}]{Kulik1977}%
  \BibitemOpen
  \bibfield  {author} {\bibinfo {author} {\bibfnamefont {I.}~\bibnamefont
  {Kulik}}\ and\ \bibinfo {author} {\bibfnamefont {A.}~\bibnamefont
  {Omelyanchuk}},\ }\bibfield  {title} {\enquote {\bibinfo {title} {Properties
  of superconducting microbridges in the pure limit},}\ }\href@noop {}
  {\bibfield  {journal} {\bibinfo  {journal} {Soviet Journal of Low Temperature
  Physics}\ }\textbf {\bibinfo {volume} {3}},\ \bibinfo {pages} {459--461}
  (\bibinfo {year} {1977})}\BibitemShut {NoStop}%
\bibitem [{\citenamefont {Linder}\ \emph {et~al.}(2008)\citenamefont {Linder},
  \citenamefont {Yokoyama}, \citenamefont {{Huertas-Hernando}},\ and\
  \citenamefont {Sudb\o}}]{Linder2008}%
  \BibitemOpen
  \bibfield  {author} {\bibinfo {author} {\bibfnamefont {J.}~\bibnamefont
  {Linder}}, \bibinfo {author} {\bibfnamefont {T.}~\bibnamefont {Yokoyama}},
  \bibinfo {author} {\bibfnamefont {D.}~\bibnamefont {{Huertas-Hernando}}}, \
  and\ \bibinfo {author} {\bibfnamefont {A.}~\bibnamefont {Sudb\o{}}},\
  }\bibfield  {title} {\enquote {\bibinfo {title} {Supercurrent switch in
  graphene {$\pi$} junctions},}\ }\href {\doibase 10/c69mcs} {\bibfield
  {journal} {\bibinfo  {journal} {Physical Review Letters}\ }\textbf {\bibinfo
  {volume} {100}},\ \bibinfo {pages} {187004} (\bibinfo {year}
  {2008})}\BibitemShut {NoStop}%
\bibitem [{Note1()}]{Note1}%
  \BibitemOpen
  \bibinfo {note} {Note that the term \protect \textit {inverse effect} cannot
  here be understood as the Onsager reciprocal proper, as our calculations are
  carried out is \protect \textit {equilibrium}.}\BibitemShut {Stop}%
\bibitem [{\citenamefont {Golubov}\ \emph {et~al.}(2004)\citenamefont
  {Golubov}, \citenamefont {Kupriyanov},\ and\ \citenamefont
  {Il'ichev}}]{Golubov2004}%
  \BibitemOpen
  \bibfield  {author} {\bibinfo {author} {\bibfnamefont {A.~A.}\ \bibnamefont
  {Golubov}}, \bibinfo {author} {\bibfnamefont {M.~Yu.}\ \bibnamefont
  {Kupriyanov}}, \ and\ \bibinfo {author} {\bibfnamefont {E.}~\bibnamefont
  {Il'ichev}},\ }\bibfield  {title} {\enquote {\bibinfo {title} {The
  current-phase relation in {{Josephson}} junctions},}\ }\href {\doibase
  10/bs7wmz} {\bibfield  {journal} {\bibinfo  {journal} {Reviews of Modern
  Physics}\ }\textbf {\bibinfo {volume} {76}},\ \bibinfo {pages} {411--469}
  (\bibinfo {year} {2004})}\BibitemShut {NoStop}%
\bibitem [{\citenamefont {Ouassou}\ and\ \citenamefont
  {Linder}()}]{Ouassou2018}%
  \BibitemOpen
  \bibfield  {author} {\bibinfo {author} {\bibfnamefont {J.~A.}\ \bibnamefont
  {Ouassou}}\ and\ \bibinfo {author} {\bibfnamefont {J.}~\bibnamefont
  {Linder}},\ }\bibfield  {title} {\enquote {\bibinfo {title} {Voltage control
  of superconducting exchange interaction and anomalous {{Josephson}}
  effect},}\ }\href@noop {} {\ }\Eprint {http://arxiv.org/abs/1810.02820}
  {arXiv:1810.02820} \BibitemShut {NoStop}%
\bibitem [{\citenamefont {Glick}\ \emph {et~al.}(2018)\citenamefont {Glick},
  \citenamefont {Aguilar}, \citenamefont {Gougam}, \citenamefont {Niedzielski},
  \citenamefont {Gingrich}, \citenamefont {Loloee}, \citenamefont {Pratt},\
  and\ \citenamefont {Birge}}]{Glick2018}%
  \BibitemOpen
  \bibfield  {author} {\bibinfo {author} {\bibfnamefont {J.~A.}\ \bibnamefont
  {Glick}}, \bibinfo {author} {\bibfnamefont {V.}~\bibnamefont {Aguilar}},
  \bibinfo {author} {\bibfnamefont {A.~B.}\ \bibnamefont {Gougam}}, \bibinfo
  {author} {\bibfnamefont {B.~M.}\ \bibnamefont {Niedzielski}}, \bibinfo
  {author} {\bibfnamefont {Eric~C.}\ \bibnamefont {Gingrich}}, \bibinfo
  {author} {\bibfnamefont {R.}~\bibnamefont {Loloee}}, \bibinfo {author}
  {\bibfnamefont {W.~P.}\ \bibnamefont {Pratt}}, \ and\ \bibinfo {author}
  {\bibfnamefont {N.~O.}\ \bibnamefont {Birge}},\ }\bibfield  {title} {\enquote
  {\bibinfo {title} {Phase control in a spin-triplet {{SQUID}}},}\ }\href
  {\doibase 10/gdzzgv} {\bibfield  {journal} {\bibinfo  {journal} {Science
  Advances}\ }\textbf {\bibinfo {volume} {4}},\ \bibinfo {pages} {eaat9457}
  (\bibinfo {year} {2018})}\BibitemShut {NoStop}%
\bibitem [{\citenamefont {Ashcroft}\ and\ \citenamefont
  {Mermin}(1976)}]{Ashcroft1976}%
  \BibitemOpen
  \bibfield  {author} {\bibinfo {author} {\bibfnamefont {N.~W.}\ \bibnamefont
  {Ashcroft}}\ and\ \bibinfo {author} {\bibfnamefont {N.~D.}\ \bibnamefont
  {Mermin}},\ }\href@noop {} {\emph {\bibinfo {title} {Solid State Physics}}},\
  \bibinfo {edition} {1st}\ ed.\ (\bibinfo  {publisher} {{Harcourt College
  Publishers}},\ \bibinfo {year} {1976})\BibitemShut {NoStop}%
\bibitem [{\citenamefont {Frigeri}\ \emph {et~al.}(2004)\citenamefont
  {Frigeri}, \citenamefont {Agterberg}, \citenamefont {Koga},\ and\
  \citenamefont {Sigrist}}]{Frigeri2004}%
  \BibitemOpen
  \bibfield  {author} {\bibinfo {author} {\bibfnamefont {P.~A.}\ \bibnamefont
  {Frigeri}}, \bibinfo {author} {\bibfnamefont {D.~F.}\ \bibnamefont
  {Agterberg}}, \bibinfo {author} {\bibfnamefont {A.}~\bibnamefont {Koga}}, \
  and\ \bibinfo {author} {\bibfnamefont {M.}~\bibnamefont {Sigrist}},\
  }\bibfield  {title} {\enquote {\bibinfo {title} {Superconductivity without
  inversion symmetry: {{MnSi}} versus {{CePt$_{3}$Si}}},}\ }\href {\doibase
  10/cw8s8d} {\bibfield  {journal} {\bibinfo  {journal} {Physical Review
  Letters}\ }\textbf {\bibinfo {volume} {92}},\ \bibinfo {pages} {097001}
  (\bibinfo {year} {2004})}\BibitemShut {NoStop}%
\bibitem [{\citenamefont {Dalichaouch}\ \emph {et~al.}(1995)\citenamefont
  {Dalichaouch}, \citenamefont {{de Andrade}}, \citenamefont {Gajewski},
  \citenamefont {Chau}, \citenamefont {Visani},\ and\ \citenamefont
  {Maple}}]{Dalichaouch1995}%
  \BibitemOpen
  \bibfield  {author} {\bibinfo {author} {\bibfnamefont {Y.}~\bibnamefont
  {Dalichaouch}}, \bibinfo {author} {\bibfnamefont {M.~C.}\ \bibnamefont {{de
  Andrade}}}, \bibinfo {author} {\bibfnamefont {D.~A.}\ \bibnamefont
  {Gajewski}}, \bibinfo {author} {\bibfnamefont {R.}~\bibnamefont {Chau}},
  \bibinfo {author} {\bibfnamefont {P.}~\bibnamefont {Visani}}, \ and\ \bibinfo
  {author} {\bibfnamefont {M.~B.}\ \bibnamefont {Maple}},\ }\bibfield  {title}
  {\enquote {\bibinfo {title} {Impurity scattering and triplet
  superconductivity in {{UPt$_{3}$}}},}\ }\href {\doibase 10/bxjzqn} {\bibfield
   {journal} {\bibinfo  {journal} {Physical Review Letters}\ }\textbf {\bibinfo
  {volume} {75}},\ \bibinfo {pages} {3938--3941} (\bibinfo {year}
  {1995})}\BibitemShut {NoStop}%
\bibitem [{\citenamefont {Geibel}\ \emph {et~al.}(1994)\citenamefont {Geibel},
  \citenamefont {Schank}, \citenamefont {J\"ahrling}, \citenamefont
  {Buschinger}, \citenamefont {Grauel}, \citenamefont {L\"uhmann},
  \citenamefont {Gegenwart}, \citenamefont {Helfrich}, \citenamefont
  {Reinders},\ and\ \citenamefont {Steglich}}]{Geibel1994}%
  \BibitemOpen
  \bibfield  {author} {\bibinfo {author} {\bibfnamefont {C.}~\bibnamefont
  {Geibel}}, \bibinfo {author} {\bibfnamefont {C.}~\bibnamefont {Schank}},
  \bibinfo {author} {\bibfnamefont {F.}~\bibnamefont {J\"ahrling}}, \bibinfo
  {author} {\bibfnamefont {B.}~\bibnamefont {Buschinger}}, \bibinfo {author}
  {\bibfnamefont {A.}~\bibnamefont {Grauel}}, \bibinfo {author} {\bibfnamefont
  {T.}~\bibnamefont {L\"uhmann}}, \bibinfo {author} {\bibfnamefont
  {P.}~\bibnamefont {Gegenwart}}, \bibinfo {author} {\bibfnamefont
  {R.}~\bibnamefont {Helfrich}}, \bibinfo {author} {\bibfnamefont {P.H.P.}\
  \bibnamefont {Reinders}}, \ and\ \bibinfo {author} {\bibfnamefont
  {F.}~\bibnamefont {Steglich}},\ }\bibfield  {title} {\enquote {\bibinfo
  {title} {Doping effects on {{UPd$_{2}$Al$_{3}$}}},}\ }\href {\doibase
  10/cdbvqj} {\bibfield  {journal} {\bibinfo  {journal} {Physica B: Condensed
  Matter}\ }\textbf {\bibinfo {volume} {199\textendash{}200}},\ \bibinfo
  {pages} {128--131} (\bibinfo {year} {1994})}\BibitemShut {NoStop}%
\bibitem [{\citenamefont {Onari}\ and\ \citenamefont
  {Kontani}(2009)}]{Onari2009}%
  \BibitemOpen
  \bibfield  {author} {\bibinfo {author} {\bibfnamefont {Seiichiro}\
  \bibnamefont {Onari}}\ and\ \bibinfo {author} {\bibfnamefont {Hiroshi}\
  \bibnamefont {Kontani}},\ }\bibfield  {title} {\enquote {\bibinfo {title}
  {Violation of {{Anderson's}} theorem for the sign-reversing \emph{s}-wave
  state of iron-pnictide superconductors},}\ }\href {\doibase 10/bfpgqh}
  {\bibfield  {journal} {\bibinfo  {journal} {Physical Review Letters}\
  }\textbf {\bibinfo {volume} {103}},\ \bibinfo {pages} {177001} (\bibinfo
  {year} {2009})}\BibitemShut {NoStop}%
\bibitem [{\citenamefont {Wang}\ \emph {et~al.}(2013)\citenamefont {Wang},
  \citenamefont {Kreisel}, \citenamefont {Hirschfeld},\ and\ \citenamefont
  {Mishra}}]{Wang2013}%
  \BibitemOpen
  \bibfield  {author} {\bibinfo {author} {\bibfnamefont {Y.}~\bibnamefont
  {Wang}}, \bibinfo {author} {\bibfnamefont {A.}~\bibnamefont {Kreisel}},
  \bibinfo {author} {\bibfnamefont {P.~J.}\ \bibnamefont {Hirschfeld}}, \ and\
  \bibinfo {author} {\bibfnamefont {V.}~\bibnamefont {Mishra}},\ }\bibfield
  {title} {\enquote {\bibinfo {title} {Using controlled disorder to distinguish
  $s_\pm$ and $s_{++}$ gap structure in {Fe}-based superconductors},}\ }\href
  {\doibase 10/c38p} {\bibfield  {journal} {\bibinfo  {journal} {Physical
  Review B}\ }\textbf {\bibinfo {volume} {87}},\ \bibinfo {pages} {094504}
  (\bibinfo {year} {2013})}\BibitemShut {NoStop}%
\bibitem [{\citenamefont {Mackenzie}\ \emph {et~al.}(1998)\citenamefont
  {Mackenzie}, \citenamefont {Haselwimmer}, \citenamefont {Tyler},
  \citenamefont {Lonzarich}, \citenamefont {Mori}, \citenamefont {Nishizaki},\
  and\ \citenamefont {Maeno}}]{Mackenzie1998}%
  \BibitemOpen
  \bibfield  {author} {\bibinfo {author} {\bibfnamefont {A.~P.}\ \bibnamefont
  {Mackenzie}}, \bibinfo {author} {\bibfnamefont {R.~K.~W.}\ \bibnamefont
  {Haselwimmer}}, \bibinfo {author} {\bibfnamefont {A.~W.}\ \bibnamefont
  {Tyler}}, \bibinfo {author} {\bibfnamefont {G.~G.}\ \bibnamefont
  {Lonzarich}}, \bibinfo {author} {\bibfnamefont {Y.}~\bibnamefont {Mori}},
  \bibinfo {author} {\bibfnamefont {S.}~\bibnamefont {Nishizaki}}, \ and\
  \bibinfo {author} {\bibfnamefont {Y.}~\bibnamefont {Maeno}},\ }\bibfield
  {title} {\enquote {\bibinfo {title} {Extremely strong dependence of
  superconductivity on disorder in {{Sr$_{2}$RuO$_{4}$}}},}\ }\href {\doibase
  10/b562dq} {\bibfield  {journal} {\bibinfo  {journal} {Physical Review
  Letters}\ }\textbf {\bibinfo {volume} {80}},\ \bibinfo {pages} {161--164}
  (\bibinfo {year} {1998})}\BibitemShut {NoStop}%
\bibitem [{\citenamefont {Abrikosov}\ and\ \citenamefont
  {Gor'kov}(1961)}]{Abrikosov1961}%
  \BibitemOpen
  \bibfield  {author} {\bibinfo {author} {\bibfnamefont {A.~A.}\ \bibnamefont
  {Abrikosov}}\ and\ \bibinfo {author} {\bibfnamefont {L.~P.}\ \bibnamefont
  {Gor'kov}},\ }\bibfield  {title} {\enquote {\bibinfo {title} {Contribution to
  the theory of superconducting alloys with paramagnetic impurities},}\
  }\href@noop {} {\bibfield  {journal} {\bibinfo  {journal} {Soviet
  Physics\textemdash{}JETP}\ }\textbf {\bibinfo {volume} {12}},\ \bibinfo
  {pages} {1243} (\bibinfo {year} {1961})}\BibitemShut {NoStop}%
\end{thebibliography}
\end{document}